\documentclass[
pra, onecolumn,
 amsmath,amssymb,superscriptaddress,
longbibliography
]{revtex4-2}
\usepackage{xcolor}
\usepackage{xparse, tensor}
\usepackage{appendix}
\usepackage{graphicx}
\usepackage{dcolumn, physics}
\usepackage{bm}
\usepackage{hyperref}
\usepackage{scalerel}
\UseRawInputEncoding

\hypersetup{
 colorlinks=true,
 linkcolor=blue,
 anchorcolor = blue,
 citecolor = blue,
 filecolor = blue,
 urlcolor = blue
}

\renewcommand{\d}[2]{\ensuremath{\frac{\text{d} #1}{\text{d} #2}}}

\usepackage{bm}

\newcommand{\be}{\begin{equation}}
\newcommand{\ee}{\end{equation}}

\begin{document}

\title{Integrable quenches in the Hubbard model}
\date{\today}

\author{Colin Rylands}
\affiliation{SISSA and INFN, via Bonomea 265,  34136 Trieste ITALY}
\author{Bruno Bertini}
\affiliation{School of Physics and Astronomy, University of Nottingham, Nottingham, NG7 2RD, UK}
\affiliation{Centre for the Mathematics and Theoretical Physics of Quantum Non-Equilibrium Systems,
University of Nottingham, Nottingham, NG7 2RD, UK}
\author{Pasquale Calabrese}
\affiliation{SISSA and INFN, via Bonomea 265,  34136 Trieste ITALY}
\affiliation{International Centre for Theoretical Physics (ICTP), Strada Costiera 11, 34151 Trieste, Italy}

\begin{abstract}
We study the quench dynamics of the one-dimensional Hubbard model through the Quench Action formalism. We introduce a class of integrable initial states --- expressed as product states over two sites --- for which we can provide an exact characterisation of the late-time regime. This is achieved by finding a closed-form expression for the overlaps between our states and the Bethe ansatz eigenstates, which we check explicitly in the limits of low densities and infinite repulsion. Our solution gives access to the stationary values attained by local observables (we show the explicit example of the density of doubly occupied sites) and the asymptotic entanglement dynamics directly in the thermodynamic limit. Interestingly, we find that for intermediate interaction strength R\'enyi entropies display a double-slope structure. 
\end{abstract}

\maketitle

\tableofcontents

\section{Introduction}
The study of interacting integrable quantum systems began almost a century ago with the exact solution of the Heisenberg chain by Hans Bethe~\cite{Bethe} and has since gone through periods of relative dormancy and activity.  These revivals have typically coincided with sudden advances in theoretical techniques, such as the introduction of the nested~\cite{Yang,  Gaudin,  GuanBatchelorLee,  ZhaoGuanLiu} or algebraic Bethe ansatz~\cite{KorepinBogoliubovIzergin} methods. We are currently experiencing one of these revivals,  triggered this time by remarkable innovations and discoveries in a range of experimental platforms which have the ability to simulate integrable models in the laboratory~\cite{Bloch, langen2015experimental,schemmer2019generalized,CavityQEDreview, RylandsGuoKeelingLevGalitski, bouchoule2022generalized,malvania2021generalized,vijayan2020time,greif2016site,omran2015microscopic,
schneider2012fermionic,hackerman2010anomalous}. The common feature of all such platforms is the ability to prepare specific initial states and monitor the resulting real time dynamics, sparking an intense theoretical investigation of this non-equilibrium setting~\cite{PolkovnikovReview, calabrese2016introduction, bastianello2022introduction, RylandsAndrei, MitraReview, bertini2021finite}. Many conceptual and technical breakthroughs on the theoretical side have since followed leading to understanding of thermalization~\cite{RigolPRL,VidmarRigol, essler2016quench}, entanglement~\cite{CalabreseCardy, calabrese2016quantum, AlbaCalabrese1,Calabrese2020, bertini2022growth} and the emergence of hydrodynamics~\cite{BertiniColluraDenardisFagotti,CastroDoyonYoshimura, doyon2020lecture, alba2021generalized, bertini2022bogoliubov} in such systems.  

A key technical breakthrough has been the introduction of the Quench Action method~\cite{CauxEssler, Caux}, which allows for the calculation of the long time steady state of a post quench integrable model and, in principle, also its finite time dynamics.  The Quench Action has so far been successfully employed in many different models~\cite{DeNardis,Brockmann, wouters2014quenching, pozsgay2014correlations, MestyanPozsgayTakacsWerner, bertini2016quantum, MestyanBertiniPiroliCalabrese,BertiniSchurichtEssler,BertiniTartagliaCalabrese,PiroliVernierCalabresePozsgay1,PiroliVernierCalabresePozsgay2, piroli2016multiparticle, piroli2016quantum, alba2016the, piroli2016exact, denardis2015relaxation, mestyan2017exact} and has since been adapted for studying entanglement~\cite{AlbaCalabreseRenyi1,AlbaCalabreseRenyi2,PiroliVernierColluraCalabrese,Mesty_n_2018,Lagnese2022} and other global properties of integrable systems out of equilibrium~\cite{PerfettoPiroliGambassi}. Although quite generally applicable, this method relies upon explicit knowledge of the overlap between the initial state of the system and the Bethe eigenstates. The latter is in itself a highly nontrivial task and overlap formulae have been found only for specific initial states and models~\cite{BrockmannDenardisWoutersCaux,Pozsgay2014, Brockmann2,BrockmannDenardisWoutersCaux2,Mazza,deLeeuw,RupasovYudson, Yudson, IyerAndrei, IyerGuanAndrei, LiuAndrei,RylandsAndreiWork2, GB1,GB2,GP, JiangPozsgay}. An important advance facilitating this endeavour has been the identification of a set of integrable initial states~\cite{PiroliPozsgayVernier}, inspired by similar notions in the context of integrable field theory~\cite{GhoshalZamolodichkov}, and related to integrable boundary conditions for classical vertex models~\cite{Pozsgay2013,PiroliPozsgayVernier1,PiroliPozsgayVernier2,RylandsAndreiWork2, pozsgay2019integrable}. These have been shown to lead to explicit factorized overlap formulas, perfectly suited for use in the Quench Action formalism. 
 
In this paper we study the integrable quench dynamics of the one-dimensional Hubbard model. This is the central theoretical model for understanding low-dimensional solids and allows for insight into an array of different possible phases and phase transitions, notably the Mott metal-insulator transition~\cite{Mahan, EsslerFrahmGohmannKlumper}. Despite the complexity of its nested Bethe ansatz solution~\cite{LiebWu1, LiebWu2}, many of the thermodynamic properties of this system are well understood within the Bethe ansatz framework~\cite{EsslerFrahmGohmannKlumper}. The non-equilibrium properties of the model are now also coming under increasing scrutiny.  Thus far however, the dynamics has been mainly studied by approximate means~\cite{MoeckelKehrein, IyerMondainiWillRigol,IllievskiDenardis,NozwaTsunetsugu1,NozwaTsunetsugu2,FavaWare,
EcksteinKollarWerner, schiro2010time, schiro2011quantum,queisser2014equilibration,riegger2015interaction,yin2016quench,schlunzen2017nonequilibrium,Ruggiero2021a,Ruggiero2021},  
while an exact description of genuine quantum quenches has been possible only in the limit of infinite repulsion~\cite{BertiniTartagliaCalabrese,TartagliaCalabreseBertini} due to the lack of manageable initial states. In this paper we fill this gap and present a family of initial states allowing for an exact Quench Action analysis of generic interactions. We identify such a solvable class of states by taking advantage of previous work on a class on integrable spin chains~\cite{GB1,GB2}. Then, we use the special  properties of these states to derive their overlaps with the Bethe eigenstates, which we explicitly test in the limits of infinite repulsion and low density. This puts us in a position to study the quench dynamics of the model using the standard Quench Action technology. In particular, we determine the long time steady state of the system and employ the Hellmann-Feynman theorem to calculate the density of doubly occupied sites.  We then go on to study the dynamics of entanglement in the scaling limit of large systems and times. In particular, we study the evolution of von Neumann entanglement entropy using the quasiparticle picture of Refs.~\cite{CalabreseCardy, AlbaCalabrese1} and that of R\'enyi entropies using the spacetime duality approach recently introduced in Ref.~\cite{bertini2022growth}. 
 
The remainder of the paper is structured as follows: in Section~\ref{SectionII} we introduce the model, discuss its properties and briefly review its Bethe ansatz solution and its thermodynamic description.  In Section~\ref{SectionIII} we describe the quench protocol and introduce a family of integrable initial states for the model.  In the subsequent section we derive an overlap formula first at infinite repulsion and then for low densities. We then use these to conjecture the general formula using a method introduced by Gombor and Bajnok~\cite{GB1,GB2}.  In Section~\ref{SectionV} we review the Quench Action formalism and apply it to the infinite repulsion and finite interaction cases. We study the properties of the long time steady state via the density of doubly occupied states in Section~\ref{SectionVIII} through the use of the Hellman-Feynmann theorem. In the penultimate section we study the dynamics of R\'enyi entropies at finite time. In the last section we summarize our work and draw our conclusions. 

\section{Hubbard Model}
\label{SectionII}

The Hubbard model describes a system of interacting spinful fermions on the lattice. Its Hamiltonian reads as  
\begin{eqnarray}
\label{HubbardHamiltonian}\nonumber
H=\sum_{j=1}^L\sum_{a=\uparrow,\downarrow}-\mathfrak{t}\left( c^\dag_{j a}c^{\phantom{\dag}}_{j+1 a}+c^\dag_{j+1 a}c^{\phantom{\dag}}_{j a}\right)+\mathfrak{u} \sum_{j=1}^Lc^\dag_{j\uparrow}c^{\phantom{\dag}}_{j\uparrow} c^\dag_{j\downarrow}c^{\phantom{\dag}}_{j\downarrow}\\
-\mu\sum_{j=1}^L\left(c^\dag_{j\uparrow}c^{\phantom{\dag}}_{j\uparrow}+c^\dag_{j\downarrow}c^{\phantom{\dag}}_{j\downarrow}\right)-h\sum_{j=1}^L\left(c^\dag_{j\uparrow}c^{\phantom{\dag}}_{j\uparrow}-c^\dag_{j\downarrow}c^{\phantom{\dag}}_{j\downarrow}\right),
\end{eqnarray}
where $c^\dag_{ja},c^{\phantom{\dag}}_{ja}$ are creation and annihilation operators for fermions at site $j$ with spin $a=\uparrow,\downarrow$ satisfying $\{c^{\phantom{\dag}}_{ja},c^{\phantom{\dag}}_{ib}\}=\{c^\dag_{ja},c^\dag_{ib}\}=0$ and $\{c^\dag_{ja},c^{\phantom{\dag}}_{ib}\}=\delta_{i,j}\delta_{a,b}$. Moreover, $L$ is the length of the chain, which we take to be even, $\mathfrak{t}$ is the hopping parameter and $\mathfrak{u}$ is the interaction strength. We have also included a chemical potential,  $\mu$, and a magnetic field, $h$, which couple to the total particle number operator
\begin{eqnarray}
\hat{N}=\hat{N}_\uparrow+\hat{N}_\downarrow=\sum_{j=1}^L\left(c^\dag_{j\uparrow}c^{\phantom{\dag}}_{j\uparrow}+c^\dag_{j\downarrow}c^{\phantom{\dag}}_{j\downarrow}\right)\,,
\end{eqnarray}
 and the $z$ component of the spin operator
\begin{eqnarray}
\hat{S}^z=\frac{1}{2}\left(\hat{N}_\uparrow-\hat{N}_\downarrow\right)=\frac{1}{2}\sum_{j=1}^L\left(c^\dag_{j\uparrow}c^{\phantom{\dag}}_{j\uparrow}-c^\dag_{j\downarrow}c^{\phantom{\dag}}_{j\downarrow}\right)\,,
\end{eqnarray}
respectively.  Both of these commute with the Hamiltonian,  $[H,\hat{N}]=[H,\hat{S}^z]=0$ and we denote the number of particles by $N$ and the total $z$-spin by $S^z= N/2-M$ where $M$ is the number of down spin particles.  

The model enjoys a number of transformations which map the Hamiltonian to itself in different regimes of the parameters. For example, we may perform a particle hole transformation,
\begin{eqnarray}\label{PHTrans}
c^\dag_{ja}\to (-1)^j c^{\phantom{\dag}}_{ja},~~c^{\phantom{\dag}}_{ja} \to(-1)^j c^\dag_{ja},
\end{eqnarray}
under which the Hamiltonian is mapped to itself, up to a constant, but with a different chemical potential and magnetic field
\begin{eqnarray}
H\left({\mathfrak{u},\mu,h}\right)\to H\left({\mathfrak{u},\mathfrak{u}-\mu,-h}\right)+\mathfrak{u}L,
\end{eqnarray}
along with $\hat{N}_{\uparrow,\downarrow}\to L-\hat{N}_{\uparrow,\downarrow}$.  Alternatively one may perform the particle-hole transformation on only one species,  say the down spins 
\begin{eqnarray}\label{ShibaTrans}
c^\dag_{j\downarrow}\to (-1)^j c^{\phantom{\dag}}_{j\downarrow},~~c^{\phantom{\dag}}_{j\downarrow} \to(-1)^j c^\dag_{j\downarrow},
\end{eqnarray}
while keeping the up spins invariant.  This Shiba transformation~\cite{EsslerFrahmGohmannKlumper}, as it is known, maps the Hamiltonian between the repulsive and attractive regimes,
\begin{eqnarray}
H\left({\mathfrak{u},\mu,h}\right)\to H\left({-\mathfrak{u},-h-\mathfrak{u}/2,\mathfrak{u}/2-\mu}\right)-(\mu+h)L,
\end{eqnarray}
along with $N_\downarrow\to L-N_\downarrow$.  Using these transformations it is sufficient to consider the model only for $N\leq L$, $2M\leq N$  and $\mathfrak{u}\geq 0$ from which the properties in the other parameter regimes can be inferred. 

In the absence of the magnetic field,  $h=0$ the  conservation of $S^z$ is enhanced to the full $SU(2)$ spin symmetry generated by $\hat{S}^z$ and the spin raising and lowering operators
\begin{eqnarray}
\hat{S}^+=\sum_{j=1}^Lc^\dag_{j\uparrow}c^{\phantom{\dag}}_{j\downarrow}, \qquad \hat{S}^-=\sum_{j=1}^Lc^\dag_{j\downarrow}c^{\phantom{\dag}}_{j\uparrow}.
\end{eqnarray}
In addition,  when the chemical potential takes the particle hole symmetric value, $\mu=\mathfrak{u}/2$ there exists an additional $SU(2)$ charge symmetry known as $\eta$ symmetry.  The latter is generated by
\begin{eqnarray}
\eta^z=\frac{1}{2}\left(L-\hat{N}\right),\qquad \eta^-=\sum_{j=1}^L(-1)^jc^\dag_{j\uparrow}c^\dag_{j\downarrow},\qquad \eta^+=\sum_{j=1}^L(-1)^jc^{\phantom{\dag}}_{j\downarrow}c^{\phantom{\dag}}_{j\uparrow},
\end{eqnarray}
which can be shown to obey ${\rm su}(2)$ commutation relations.  At generic values of the chemical potential and magnetic field we have 
\begin{equation}
[H,\eta^\pm]=\mp(\mathfrak{u}-2\mu)\eta^\pm,\qquad [H,\hat{S}^\pm]=\mp 2h \hat{S}^\pm,
\label{eq:HSHeta}
\end{equation}
meaning that the eigenstates of the model arrange themselves into $SU(2)$ multiplets. 

The Hamiltonian is integrable and its exact solution was found by Lieb and Wu \cite{LiebWu1, LiebWu2}. The eigenstates of the model are labelled by the ``momenta" $k_j,~j=1,\dots N$ and ``spin rapidities" $\lambda_\gamma, ~\gamma=1,\dots,M$ and are given by
\begin{eqnarray}
\ket{\{k_j\}\{\lambda_\gamma\}}=\sum^L_{n_1\leq n_2\dots\leq n_N}\sum_{a_1\dots a_N=\uparrow,\downarrow}\psi(\vec{n},\vec{a})c^\dag_{n_1,a_1}\dots c^\dag_{n_N,a_N}\ket{0}\,,
\end{eqnarray}
where  $\ket{0}$ is the empty chain i.e. $c^{\phantom{\dag}}_{ja}\ket{0}=0$ and the unnormalised  wavefunction reads as
\begin{eqnarray}
\psi(\vec{n},\vec{a})&=&\sum_{P\in \mathcal{S}_N}(-1)^Pe^{i\sum_j k_{P_j}n_j}\varphi_P(\vec{a}),\\
\varphi_P(\vec{a})&=&\sum_{Q\in \mathcal{S}_M}\prod_{1\leq Q_\gamma <Q_\beta\leq M}\frac{\lambda_{Q_\gamma}-\lambda_{Q_\beta}-i U}{\lambda_{Q_\gamma}-\lambda_{Q_\beta}}\prod^M_{l=1}F_P(\lambda_{Q_l},y_l)\,,\\
F_{P}(\lambda_{Q_l},y_l)&=&\frac{iU}{\lambda_{Q_l}-s_{P_{y_l}}+i U/2}\prod_{j=1}^{y_l-1}\frac{\lambda_{Q_l}-s_{P_{j}}-i U/2}{\lambda_{Q_l}-s_{P_{j}}+i U/2}\,.
\end{eqnarray}
Here $\mathcal S_N$ denotes the group of permutations of $N$ elements, $\varphi_P(\vec{a})$ is the spin part of the wavefunction and $y_l$ are the positions of the $M$ up spins amongst the $N$ particles i.e. $y_j=l$ means that the $l^\text{th}$ particle is the $j^\text{th}$ up spin.  We have introduced the reduced  interaction strength in terms of which the results are expressed,
\begin{eqnarray}
U=\frac{\mathfrak{u}}{2\mathfrak{t}},
\end{eqnarray}
and used the shorthand notation
\be
s_{k_j}\equiv \sin{(k_j)}, \qquad c^{\phantom{\dag}}_{k_j}\equiv\cos(k_j).
\label{eq:shorthand}
\ee  
The momenta and spin raipidities are not free but are quantized and coupled to one another through the nested Bethe Ansatz equations 
\begin{eqnarray}\label{BAE1}
e^{-i k_jL }&=&\prod_{\gamma}^M\frac{\lambda_\gamma-s_{k_j}-iU/2}{\lambda_\gamma-s_{k_j}+iU/2},\\\label{BAE2}
\prod_{i=1}^N\frac{\lambda_\gamma-s_{k_i}-iU/2}{\lambda_\gamma-s_{k_i}+iU/2}&=&\prod_{\beta\neq\gamma}^M\frac{\lambda_\gamma-\lambda_\beta-iU}{\lambda_\gamma-\lambda_\beta+iU}.
\end{eqnarray}
The energy and total momentum of this state are 
\begin{eqnarray}
\label{eq:energy}
E=-2\mathfrak{t}\sum_{j=1}^N\cos{(k_j)}-\mu N-h(N-2M),~~P=\sum_{j=1}^Nk_j~\text{mod} ~2\pi.
\end{eqnarray} 
The norm of the Hubbard Bethe states was found by Gohman and Korepin~\cite{GK} and is given by 
\begin{eqnarray}\label{Norm}
\braket{\{k_j\}\{\lambda_\gamma\}}{\{k_j\}\{\lambda_\gamma\}}=U^M\prod_{j=1}^N\cos{(k_j)}\prod_{1\leq\gamma<\beta \leq M}\left[\frac{(\lambda_\gamma-\lambda_\beta)^2+U^2}{(\lambda_\gamma-\lambda_\beta)^2}\right]\text{det}[G].
\end{eqnarray}
Here $G$ is a  $(N+M)\times (N+M)$ matrix of Gaudin type which typically appears in the norm of Bethe ansatz wavefunctions.  It's components are
\begin{eqnarray} \label{GaudinMatrix}
G_{i,j}&=&\delta_{ij}\left[\frac{L}{\cos{k_j}}+\sum_\gamma^M\phi_1(\lambda_\gamma-s_{k_j})\right],\\
G_{i,N+\gamma}&=&G_{N+\gamma,i}=-\phi_1(\lambda_\gamma-s_{k_j}),\\
 G_{N+\gamma,N+\beta}&=&\delta_{\gamma \beta}\left[\sum_{j=1}^N\phi_1(\lambda_\gamma-s_{k_j})-\sum_{\gamma=1}^M\phi_2(\lambda_\gamma-\lambda_\gamma)\right]+\phi_2(\lambda_\gamma-\lambda_\beta),
\end{eqnarray}
for $1\leq i,j\leq N$,~ $1\leq \gamma,\beta\leq M$ and with 
\begin{eqnarray}\label{phin}
\phi_n(x)=\frac{nU}{(nU/2)^2+x^2}.
\end{eqnarray} 
The states described thus far are only the highest weight states of $SU(2)$ spin and $\eta$ symmetry multiplets.  The remainder of the states are obtained by acting on these with the lowering operators $\hat{S}^-$ and $\eta^-$.  In this work we shall only need the descendant state obtained through application of $\eta^-$ which we denote
\begin{eqnarray}\label{descendant}
\ket{\{k_j\}\{\lambda_\gamma\};m}=\sqrt{\frac{(L-N-m)!}{m!(L-N)!}}(\eta^-)^m\ket{\{k_j\}\{\lambda_\gamma\}}\,,
\end{eqnarray}
where $0\leq m\leq L-N$. The prefactor included in the definition ensures that the norm of this state is the same as the highest weight state  i.e.  
\begin{equation}
\braket{\{k_j\}\{\lambda_\gamma\};m}{\{k_j\}\{\lambda_\gamma\};m}=\braket{\{k_j\}\{\lambda_\gamma\}}{\{k_j\}\{\lambda_\gamma\}}.
\end{equation}
Using the commutation relations \eqref{eq:HSHeta} the energy of the descendant states is easily found to be 
\begin{equation}
H\ket{\{k_j\}\{\lambda_\gamma\};m}=[E+m(\mathfrak{u}-2\mu)]\ket{\{k_j\}\{\lambda_\gamma\};m}.
\end{equation}

\subsection{String Hypothesis and Thermodynamic limit}
\label{sec:TBA}

As is typically the case, the Bethe ansatz equations~\eqref{BAE1},~\eqref{BAE2} admit both real and complex solutions. In thermodynamic limit
\be
\lim_{\rm th}\equiv \lim_{\substack{N,L,M\to\infty\\ N/L = \text{fixed} \\ M/L = \text{fixed}}},
\ee
however, the system can be described assuming that the complex solutions form regular patterns in the complex plane known as strings~\cite{Takahashi,EsslerKorepinSchoutens}. In the Hubbard model the strings come in two types:
\begin{itemize}
\item[(i)] ``$\lambda$ strings" in which the $\lambda$ spin rapidities take the form
\begin{eqnarray}\label{lstrings}
\lambda_{n,j,\gamma}=\lambda_{n,\gamma}+i(n+1-2j)U/2, \qquad \lambda_{n,\gamma}\in\mathbb R, \qquad j=1,\dots, n,
\end{eqnarray}
up to a correction vanishing exponentially in the thermodynamic limit.  A $\lambda$ string describes a multiparticle excitation in the spin sector of the model.  It has $S^z=-n$ but zero charge, $N=0$ while its energy and momentum are also zero.  
\item[(ii)] ``$k\!-\!\lambda$ strings" consisting of a set of $2n$ complex momenta $k_{l,\gamma},~l=1,\dots, 2n$  and a $\lambda$ string. The latter  takes same  form as above 
\begin{equation}
\lambda'_{n,j,\gamma}=\lambda'_{n,\gamma}+i(n+1-2j)U/2, \qquad \lambda'_{n,\gamma}\in\mathbb R,\qquad j=1,\dots, n,
\label{eq:klstring1}
\end{equation}
whereas the former are given by
\begin{eqnarray}\label{klstring}
k_{1,\gamma}&=&\pi-\arcsin{(\lambda'_{n,\gamma}+ni U/2)}\\\nonumber
k_{2,\gamma}&=&\arcsin{(\lambda'_{n,\gamma}+(n-2)i U/2)}\\\nonumber
k_{3,\gamma}&=&\pi-k_{2,\gamma} \\\nonumber
&&\vdots\\\nonumber
k_{2n-1,\gamma}&=&\pi-k_{2n-2,\gamma}\\\nonumber
k_{2n,\gamma}&=&\pi-\arcsin{(\lambda'_{n,\gamma}-i nU/2)}.
\end{eqnarray}
A $k\!-\!\lambda$ string describes a multiparticle excitation, this time in the charge sector of the model.  It has charge $N=2n$ but is a spin singlet $S^z=0$ and its energy and momentum are
\begin{eqnarray}\label{KLEnergy}
e_{k-\lambda, n}(\lambda')&=&-2\mathfrak{t}\sum_{j=1}^{2n}\cos(k_{j})=4\mathfrak{t}\mathfrak{R}\left[\sqrt{1-(\lambda'+ni U/2})^2\right]\notag\\
&=&2Un+\mathfrak{t}\int_{-\pi}^\pi\frac{{\rm d}k}{\pi}\cos^2{(k)}\phi_n(\lambda'-\sin(k)),\label{IntEnergy}\\
p_{k-\lambda, n}(\lambda')&=&\sum_{j=1}^{2n}k_{j}=-2\mathfrak{R}\left[\arcsin{(\lambda'+ni U/2)}\right]~\text{mod} 2\pi\notag\\
&=&\int_{-\pi}^\pi\frac{{\rm d}k}{\pi}\arctan\left(\frac{2(\sin(k)-\lambda')}{nU}\right)~\text{mod} 2\pi.
\label{IntMomentum}
\end{eqnarray}
\end{itemize}
We note that the energy and momentum reported above are the energy increase caused by the addition of a string excitation to a state with no other particles and hence they are referred to as \emph{bare}. When other particles are also present  the energy increase caused by the addition of an excitation is different. This fact is typically expressed by saying that energy and momentum of an excitation are ``dressed by the interactions". 

Under the string hypothesis, a given eigenstate is described by its content of $\lambda$ strings, $k-\lambda$ strings, and real momenta. The key simplification introduced by this hypothesis is that also $\lambda$ strings, $k-\lambda$ strings are fully specified by real numbers, i.e., $\lambda_{n,\gamma}$ and $\lambda'_{n,\gamma}$ (cf.~\eqref{lstrings}--\eqref{klstring}). These quantities, called \emph{string centers}, are interpreted as the real rapidities of different elementary particles. 

In any finite volume real rapidities take discrete values which are obtained solving the Bethe equations \eqref{BAE1} and \eqref{BAE2}. In the thermodynamic limit, however, they take real values and can be described by distributions. To this aim we follow Ref.~\cite{takahashi1972one} and introduce the distributions of ``particles", ``holes" $\rho(k)$ and $\rho^h(k)$. In essence these quantities keep track of the distribution of real rapidities. Analogously, we introduce $\{\sigma_n(\lambda),~\sigma_n'(\lambda)\}$, and $\{\sigma_n^h(\lambda),\sigma_n^{\prime\, h}(\lambda)\}$, specifying the distribution of the centres of $\lambda$ and $k-\lambda$ strings.

These distributions are coupled through the following set of integral equations originating from  \eqref{BAE1} and \eqref{BAE2} and known as the Bethe-Takahashi equations~\cite{takahashi1972one, EsslerFrahmGohmannKlumper}
\begin{eqnarray}\label{ContBAE1}
&&\!\!\!\!\!\!\!\!\!\rho(k)+\rho^h(k)=\frac{1}{2\pi}+\cos(k)\sum_{n=1}^\infty\int \frac{{\rm d}\lambda}{2\pi} \phi_n(s_k-\lambda)\left[\sigma_n(\lambda)+\sigma'_n(\lambda)\right],\\\label{ContBAE2}
&&\!\!\!\!\!\!\!\!\!\sigma_n(\lambda)+\sigma_n^h(\lambda)=\int_{-\pi}^\pi \frac{{\rm d}k}{2\pi}\phi_n(s_k-\lambda)\rho(k)-\sum_{m=1}^\infty T_{nm}*\sigma_m(\lambda),\\\label{ContBAE3}
&&\!\!\!\!\!\!\!\!\!\sigma'_n(\lambda)+{\sigma'}_n^h(\lambda)=\frac{1}{\pi}\mathfrak{R}\!\!\left[\frac{1}{\sqrt{1-(\lambda+inU/2)^2}}\right]\!\!-\!\!\int_{-\pi}^\pi \frac{{\rm d}k}{2\pi}\phi_n(s_k-\lambda)\rho(k)-\!\!\sum_{m=1}^\infty T_{nm}*\sigma'_m(\lambda),
\end{eqnarray}
where we used the shorthand notation \eqref{eq:shorthand}, we denoted by $*$ the convolution 
\be
f*g(x)=\int_{-\infty}^\infty {\rm d}\lambda f(x-\lambda)g(\lambda),
\ee
and we introduced  
\begin{eqnarray}\label{TMN}
T_{mn}(\lambda)&=&\frac{1}{2\pi}\phi_{m+n}(\lambda)+\frac{1}{2\pi}\phi_{|m-n|}(\lambda)+\frac{1}{\pi}\sum_{j=1}^{m-1}\phi_{m+n-2j}(\lambda),\qquad m\neq n,\\
 T_{nn}(\lambda)&=&\frac{1}{2\pi}\phi_{2n}(\lambda)+\frac{1}{\pi}\sum_{j=1}^{n-1}\phi_{2n-2j}(\lambda).
\end{eqnarray}
These expressions can be brought to a partially decoupled form which is more convenient for numerical analysis.  To do this we use
\begin{equation}
\label{eq:Tinverse}
\sum_{n}[\delta+T]_{mn}^{-1}*f_n(x) =f_m(x)-s*[f_{m+1}(x)+f_{m-1}(x)],
\end{equation}   
where 
\be
s(x)\equiv\frac{1}{2U}\text{sech}(\frac{\pi x}{U}), 
\ee
fulfils
\begin{equation}
s*[\phi_{m+1}(x)+\phi_{m-1}(x)]= \phi_{m}(x)\,,\qquad m=1,\ldots,
\end{equation}
and we set $\phi_{0}(x)=\delta(x)$. Using \eqref{eq:Tinverse} in \eqref{ContBAE2} and \eqref{ContBAE3} we arrive at
\begin{eqnarray}\label{continuumBAE}
\sigma_n(\lambda)+\sigma_n^h(\lambda) &=&s*[\sigma_{n-1}^h(\lambda)+\sigma_{n+1}^h(\lambda) ]+\delta_{n,1}\int_{-\pi}^\pi {\rm d}k\, s(s_k-\lambda)\rho(k),\\
\sigma'_n(\lambda)+\sigma_n^{\prime\,h}(\lambda) &=&s*[\sigma_{n-1}^{\prime\,h}(\lambda)+\sigma_{n+1}^{\prime\,h}(\lambda) ]-\delta_{n,1}\int_{-\pi}^\pi {\rm d}k\, s(s_k-\lambda)\left[\rho(k)-\frac{1}{2\pi}\right],\\
\rho(k)+\rho^h(k) &=&\frac{1}{2\pi}+{\cos{(k)}}\!\int_{-\infty}^\infty \!\!\!\!\!{\rm d}\lambda\! \left[\frac{\phi_1(s_k-\lambda)}{2\pi}\sigma_0(\lambda)\!-\!s(s_k-\lambda)(\sigma_1^h(\lambda)+\sigma^{\prime\,h}_1(\lambda))\right]\!\!.
\end{eqnarray}
Here we have introduced
\begin{eqnarray}
\sigma_0(\lambda)=\int_{-\pi}^\pi {\rm d}k s(\lambda-s_k)\rho_0(k),\qquad
\rho_0(k)=\frac{1}{2\pi}+\cos{(k)}\int_{-\infty}^\infty\frac{d\nu}{2\pi}\frac{J_0(\nu)\cos{(\nu s_k)}}{1+e^{U|\nu|}},
\end{eqnarray} 
where $J_0(\nu)$ is zeroth order Bessel function of the first kind.  

Considering the thermodynamic limit of \eqref{eq:energy} we have that the state described by the root densities $\{\rho(k), \sigma_n(\lambda), \sigma_n'(\lambda)\}$ has energy density given by 
\begin{eqnarray}
\lim_{\rm th}\frac{E}{L}&=&\int_{-\pi}^\pi {\rm d}k\left[-2\mathfrak{t}\cos{(k)}-h-\mu\right]\rho(k)+2h\sum_{n=1}^\infty n\int_{-\infty}^\infty {\rm d}\lambda \sigma_n(\lambda)\nonumber\\
&&+\sum_{n=1}^\infty\int_{-\infty}^\infty {\rm d}\lambda\left( 4\mathfrak{t}\mathfrak{R}\left[\sqrt{1-(\lambda+i nU/2)^2}\right]-2\mu  n\right)\sigma'_n(\lambda).
\end{eqnarray}
Moreover, the densities of the various string types read as 
\begin{eqnarray}
&&\lim_{\rm th}\frac{M_n}{L}=\int_{-\infty}^\infty {\rm d}\lambda\, \sigma_n(\lambda), \qquad\qquad\qquad \lim_{\rm th}\frac{M'_n}{L}=\int_{-\infty}^\infty {\rm d}\lambda\, \sigma'_n(\lambda),\\
&& \lim_{\rm th}\frac{N}{L}-\sum_n^\infty 2n \frac{M'_n}{L}=\int_{-\pi}^\pi {\rm d}k\, \rho(k).
\end{eqnarray}

\section{Quench Protocol }\label{SectionIII}

We study the non-equilibrium dynamics of the Hubbard model by means of a quantum quench.  In this protocol  the system is initially prepared in some state $\ket{\Psi_0}$, which is not an eigenstate of the Hamiltonian~\eqref{HubbardHamiltonian}, and is then allowed evolve according to the Schr\"odinger equation, i.e.
\be
\ket{\Psi_t}=e^{-iHt}\ket{\Psi_0}, \qquad t\geq 0\,. 
\ee
The resulting time evolution can in principle be evaluated by inserting a complete set of eigenstates of $H$, which evolve via a simple phase,  and computing the overlaps between eigenstates and the initial state. 

Our initial state shall be chosen from a family of integrable initial states of the Hubbard model which lend themselves to explicit evaluation of an overlap formula~\cite{PiroliPozsgayVernier}. To determine these states we use the results of Refs.~\cite{GB1, GB2}, where the integrable initial states of certain $SU(2|2)$ invariant spin chains are derived. Since the Hubbard model is known to arise as a limit of such spin chains~\cite{Beisert}, we can directly read integrable initial states of the Hubbard model. As in other lattice models they take a two site product form and can be expressed through a $K$ matrix,
\begin{eqnarray}\label{Kmatrix}
K(\bm{\kappa},\alpha')&=&\begin{pmatrix}
\kappa_1& \kappa_2+\alpha'& 0&0\\
\kappa_2-\alpha' & \kappa_3 &0&0\\
0&0&0& \alpha'\\
0&0&-\alpha'& 0
\end{pmatrix},
\end{eqnarray}
where $\alpha'$ is some free parameter and the remaining parameters $\bm{\kappa}=(\kappa_1,\kappa_2,\kappa_3)$ are restricted by the condition $\kappa_1\kappa_3-\kappa_2^2=1$. The integrable states are a product over two site states with $K_{mn}$ as the coefficients,
\begin{eqnarray}\label{initalstate}
\ket{\Psi_0}=\otimes_{l=1}^{L/2}\ket{K}_l,\qquad\ket{K}_l=\sum_{m,n} K_{mn}\ket{m}_{2l-1}\otimes \ket{n}_{2l}.
\end{eqnarray}
Here the single site states are bosonic $\ket{1}_{l}=\ket{0}_l,~\ket{2}_l=c^\dag_{l,\uparrow}c^\dag_{l,\downarrow}\ket{0}=\ket{\uparrow\downarrow}$ or fermionic $\ket{3}_l=c^\dag_{l,\uparrow}\ket{0}=\ket{\uparrow},~\ket{4}_l=c^\dag_{l,\downarrow}\ket{0}=\ket{\downarrow}$. More explicitly the integrable initial states of the Hubbard model are a product over
\begin{eqnarray}\label{Kstate}
\ket{K}_l&=&\left[\kappa_1\ket{0}_{2l-1}\ket{0}_{2l}+\kappa_3\ket{\uparrow\downarrow}_{2l-1}\ket{\uparrow\downarrow}_{2l}\right]\\\nonumber
&&+\left[(\kappa_2+\alpha')\ket{\uparrow\downarrow}_{2l-1}\ket{0}_{2l}+(\kappa_2-\alpha')\ket{0}_{2l-1}\ket{\uparrow\downarrow}_{2l}\right]\notag\\
&&+\alpha'\left[\ket{\uparrow}_{2l-1}\ket{\downarrow}_{2l}-\ket{\downarrow}_{2l-1}\ket{\uparrow}_{2l}\right]\,.
\end{eqnarray}
A second type of initial state was also identified in~\cite{GB1}. In the Hubbard limit this can be shown to be related to the above state by a Shiba transformation~\eqref{ShibaTrans} and by combining this property with the transformations of the Hamiltonian~\eqref{PHTrans},\eqref{ShibaTrans} the resulting quench dynamics can be understood by considering~\eqref{Kmatrix}.  

For simplicity we restrict ourselves to the case where the spin structure of the state is the same as that of a dimer and $N\leq L$ by taking the limit 
\begin{equation}
\kappa_1\to\infty,\qquad \kappa_2\to0,\qquad  \alpha'\to\infty, \qquad \alpha'/\kappa_1\equiv \alpha={\rm fixed}\,.
\end{equation}
In this case we can write the initial state in the form
\begin{eqnarray}
\label{eq:ourinitialstate}
\ket{\Psi_0}&=&\mathcal{N}\exp\left[\sum_{l=1}^{L/2}\alpha X_l-\frac{1}{2}\alpha^2 X^2_l\right]\ket{0},\\
X_l&=&c^{\dag}_{2l-1\uparrow}c^{\dag}_{2l-1\downarrow}-c^{\dag}_{2l\uparrow}c^{\dag}_{2l\downarrow}+c^{\dag}_{2l-1\uparrow}c^{\dag}_{2l\downarrow}-c^{\dag}_{2l-1\downarrow}c^{\dag}_{2l\uparrow},\notag\\
X_l^2&=&-4c^{\dag}_{2l-1\uparrow}c^{\dag}_{2l-1\downarrow}c^{\dag}_{2l\uparrow}c^{\dag}_{2l-1\downarrow},\notag\\
X_l^3&=&0,\notag
\end{eqnarray}
with the norm given by $\mathcal{N}=\left[1+4\alpha^2\right]^{L/4}$.  Note this state is not an eigenstate of particle number but does become one in the limit $\alpha\to\infty$ in which case it is half filled $N=L$.   
The average particle number can be calculated straightforwardly 
\begin{eqnarray}\label{AverPart}
\frac{1}{L} \matrixel{\Psi_0}{\hat{N}}{\Psi_0}=\frac{4\alpha^2}{1+4\alpha^2}
\end{eqnarray}  
which is conserved by the dynamics. Moreover, we note that $\ket{\Psi_0}$ is also not an eigenstate of the one-site shift operator. To facilitate the Quench Action treatment we therefore consider its translational invariant version 
\be
|{\tilde \Psi_0}\rangle \longmapsto \frac{1-T}{\sqrt 2} \ket{\Psi_0}\,, 
\ee
where $T$ is the one-site shift operator. This replacement is harmless as long as there are no conserved charges not invariant under a one-site shift and, therefore, there is a restoration of one-site shift invariance in the stationary state.  Finally, unless we explicitly state otherwise, from now on we set $h=\mu=0$ in the time evolving Hamiltonian. 



\section{Overlap formula}\label{SectionIV}
We want to calculate the squared overlap between the initial state and a Bethe state,
\begin{eqnarray}
\frac{|\braket{\Psi_0}{\{k_j\}\{\lambda_\gamma\};n}|^2}{\|\ket{\{k_j\}\{\lambda_\gamma\};n}\|^2},
\end{eqnarray}
and furthermore to obtain a convenient expression which will allow us to take the thermodynamic limit and facilitate our study of the quench dynamics.  Our choice of initial states will allow us to achieve this goal which we shall tackle in two stages. First,  we derive the exact overlap in the infinite repulsion limit $U\to \infty$. Then, we consider the finite $U$ case and derive a formula within the low density limit, $N\ll L$.  Finally we use this expression to conjecture a formula valid at arbitrary filling following the method of Gombor \& Bajnok~\cite{GB1,GB2}. 

\subsection{Infinite $U$ overlap}
\label{sec:infiniteUover}

In the infinite-repulsion limit the system simplifies significantly resulting in a decoupling of the charge and spin degrees of freedom,  while at the same time projecting out certain parts of the Hilbert space,  namely states with doubly occupied states. Note that in this limit the model does not have the $SU(2)$ $\eta$ symmetry but retains the full spin symmetry of the model while the particle-hole and Shiba transformations also remain valid. 

In order to take the limit it is necessary to rexamine the Bethe equations. Therein we see that we can rescale the spin rapidities as $\lambda\to U\lambda$ and then take the limit, in which case  
\begin{eqnarray}\label{inftyUBAE1}
e^{-i k_jL }&=&\prod_{\gamma}^M\frac{\lambda_\gamma-i/2}{\lambda_\gamma+i/2},\\\label{inftyUBAE2}
\left[\frac{\lambda_\gamma-i/2}{\lambda_\gamma+i/2}\right]^N&=&\prod_{\beta\neq\gamma}^M\frac{\lambda_\beta-\lambda_\gamma-i}{\lambda_\beta-\lambda_\gamma+i},
\end{eqnarray}
resulting in a partial decoupling of the spin and charge degrees of freedom.  The latter becoming equivalent to a homogeneous $N$-site XXX chain and the former describing free fermions modulo a relationship between the momentum of the fermions and the total momentum of the spin chain, the right hand side of~\eqref{inftyUBAE1}.

The eigenstates also decouple and become a product of a Slater determinant for the charge degrees of freedom times the spin part $\varphi_{XXX}(\vec{a})$ which is the wavefunction for the homogeneous $XXX$ chain
\begin{eqnarray}
\psi(\vec{n},\vec{a})=\text{det}[e^{ik_jn_l}]\varphi_{XXX}(\vec{a}).
\end{eqnarray}
The norm of the state also factorizes
\begin{eqnarray}
\braket{\{k_j\}\{\lambda_\gamma\}}{\{k_j\}\{\lambda_\gamma\}}=L^N\prod_{1\leq\gamma<\beta \leq M}\left[\frac{(\lambda_\gamma-\lambda_\beta)^2+U^2}{(\lambda_\gamma-\lambda_\beta)^2}\right]\text{det}[G_{XXX}].
\end{eqnarray}

In order for us to perform checks on our calculations in this limit it is convenient to project the initial state into the same Hilbert space as the infinite $U$ Hubbard by projecting out the doubly occupied sites which are forbidden in this limit. After doing this the initial state takes the form
\begin{eqnarray}
\ket{\Psi^\infty_0}=\left[\frac{1}{1+2\alpha^2}\right]^{L/4}\exp\left[\alpha\sum_{l=1}^{L/2} c^\dag_{2l-1\downarrow}c^\dag_{2l\uparrow}- c^\dag_{2l-1\uparrow}c^\dag_{2l\downarrow}\right]\ket{0}.
\end{eqnarray}
Note here the change in normalization of the state. The overlap between the initial state and a normalized $N$ particle state is easily written out
\begin{eqnarray}\nonumber
\frac{\braket{\Psi_0^\infty}{\{k_k\},\{\lambda_\gamma\}}}{\sqrt{\braket{\{k_j\}\{\lambda_\gamma\}}{\{k_j\}\{\lambda_\gamma\}}}}&=&\frac{(\alpha/L)^{N/2}}{\left(1+2\alpha^2\right)^{L/4}}\sum_{l_1<\dots< l_{N/2}}\sum_{n_1<\dots<n_N}\text{det}[e^{ik_jn_l}]\\\label{inftyoverlap}
&&\qquad\qquad \times\prod_{j=1}^{N/2}\delta_{n_{2j-1},2l_j-1}\delta_{n_{2j},2l_j} \left[\frac{\varphi^\text{d}_{XXX}}{\|\varphi\|}\right],
\end{eqnarray}
where ${\varphi^\text{d}_{XXX}}/{\|\varphi\|}$ is the normalized overlap between a dimer  initial state and an eigenstate of the homogeneous $XXX$ model
\begin{eqnarray}
\varphi^\text{d}_{XXX}\equiv \sum_{\vec{a}}\prod_{j=1}^{N/2}\epsilon^{2j-1, 2j}\varphi_{XXX}(\vec{a}),
\end{eqnarray}
with $\epsilon^{i i+1}$ denoting the 2d Levi--Civita tensor on the spin space of $i$ and $i+1$ particles.  Since this quantity has been calculated in \cite{Pozsgay2014}, we focus on the other terms first. We reduce them to a convenient form by first expressing the Slater determinant above as a sum over permutations in $\mathcal{S}_N$. Explicitly we have 
\begin{eqnarray}
&&\sum_{l_1<\dots< l_{N/2}}\sum_{n_1<\dots<n_N}\text{det}[e^{ik_jn_l}]\prod_{j=1}^{N/2}\delta_{n_{2j-1},2l_j-1}\delta_{n_{2j},2l_j}\notag\\
&=& \sum_{P\in \mathcal{S}_N}(-1)^P\sum_{l_1<\dots< l_{N/2}}\sum_{n_1<\dots<n_N}e^{i\sum_{j=1}^Nk_{P_l}n_l}\prod_{j=1}^{N/2}\delta_{n_{2j-1},2l_j-1}\delta_{n_{2j},2l_j}\notag\\
&=&\sum_{P\in \mathcal{S}_N}(-1)^P\sum_{l_1<\dots< l_{N/2}}e^{2i\sum_{l=1}^{N/2}(k_{P_{2j-1}}+k_{P_{2j}})l_j}\prod_{j=1}^{N/2}\left[\frac{e^{-ik_{P_{2j-1}}}-e^{-ik_{P_{2j}}}}{2}\right].
\end{eqnarray}
In going to the last line we make repeated use of the identity 
\begin{eqnarray}
\sum_{P\in \mathcal{S}_N} f(P_1,...,P_{N}) = \frac{1}{2}\sum_{P\in \mathcal{S}_N} f(P) + \frac{1}{2}\sum_{P'\in \mathcal{S}_N} f(P')
\end{eqnarray}
where $P'$ differs from $P$ only because of the exchange of $P_{2j}$ and $P_{2j-1}$. We then replace the ordered sum over two site blocks, $l_j$ with an unordered sum by combining it with the sum over $P$ to get 
\begin{eqnarray}
\sum_{l_1<\dots< l_{N/2}}\sum_{n_1<\dots<n_N}\text{det}[e^{ik_jn_l}]\prod_{j=1}^{N/2}\delta_{n_{2j-1},2l_j-1}\delta_{n_{2j},2l_j}= L^{N/2}\prod_{j=1}^{n}\left[-i \sin{(k_j)}\right]\prod_{i=1}^{n^\pi}\left[\cos{(k_l)}\right].
\label{InftyChargeOver}
\end{eqnarray}
Where we use the fact that the sum over $l_j$ is only non zero if the momenta form pairs which are either $\{k_j,-k_j\}$ or $\{k_l,\pi-k_l\}$.  We  take there to be $n$ of the former pairs and $n^\pi$ of the latter with $n+n^\pi=N/2$.  

For the spin sector it has been shown that a dimer initial state has non zero overlap only with parity invariant states of the $XXX$ model whose rapidities have a pair structure $\{\lambda_\gamma,-\lambda_\gamma\}$  with $\lambda_\gamma>0$ and moreover that the total number of rapidities should be half the length of the chain, i.e.  $M=N/2$.  Combining the explicit result for the $XXX$ dimer with \eqref{InftyChargeOver} we arrive at
\begin{eqnarray}\label{InftyOverlap}
\frac{|\braket{\Psi_0^\infty}{\{k_j\}\{\lambda_\gamma,-\lambda_\gamma\}}|^2}{\|\ket{\{k_j\}\{\lambda_\gamma,-\lambda_\gamma\}}\|^2}=\frac{\alpha^{N}}{\left(1+2\alpha^2\right)^{L/2}}\prod_{j=1}^{n}\sin^2{(k_j)}\prod_{i=1}^{n^\pi}\cos^2{(k_l)}\prod_{\gamma}^{N/4}\frac{1}{\lambda_\gamma^2(\lambda_\gamma^2+1/4)}\frac{\text{det}[G^+_{XXX}]}{\text{det}[G^-_{XXX}]}. \quad
\end{eqnarray}
In the last term we have a ratio of Gaudin determinants $G^\pm_{XXX}$ which is typical of overlap formulae for integrable initial states.  The components of the matrices are 
\begin{eqnarray}\label{InftyGaudinMatrix}
[G^{\pm}_{XXX}]_{\gamma \beta}=\delta_{\gamma \beta}\left[N\bar{\phi}_1(\lambda_\gamma)-\sum_{\gamma=1}^{N/4}\bar{\phi}^\pm_2(\lambda_\gamma,\lambda_\gamma)\right]+\bar{\phi}^\pm_2(\lambda_\gamma,\lambda_\beta),\\
\bar{\phi}^\pm_n(\lambda,\mu)=\bar{\phi}_n(\lambda-\mu)\pm \bar{\phi}_n(\lambda+\mu),~\bar{\phi}_n(\lambda)=\frac{n}{\lambda^2+(n/2)^2},
\end{eqnarray}
As shown in Refs.~\cite{BrockmannDenardisWoutersCaux, MestyanPozsgayTakacsWerner} we have 
\be
\lim_{\rm th} \frac{\text{det}[G^+_{XXX}]}{\text{det}[G^-_{XXX}]}= 1.
\ee 

We can prove that Eq.~\eqref{InftyOverlap} exhausts the set of states having non zero overlap with $\ket{\Psi_0^\infty}$ by considering its normalization and inserting a complete set of states
\begin{eqnarray}\nonumber
\braket{\Psi_0^\infty}{\Psi_0^\infty}&=&\sum_{N=0}^L\sum_{\text{states}}\frac{|\braket{\Psi_0^\infty}{\{k_j\}\{\lambda_\gamma,-\lambda_\gamma\}}|^2}{\braket{\{k_j\}\{\lambda_\gamma,-\lambda_\gamma\}}{\{k_j\}\{\lambda_\gamma,-\lambda_\gamma\}}}\\\nonumber
&=&\sum_{m}^{L/2}\frac{\alpha^{2m}}{\left(1+2\alpha^2\right)^{L/2}}\sum_{\text{states}}\prod_{j=1}^{n}\sin^2{(k_j)}\prod_{i=1}^{n^\pi}\cos^2{(k_l)}\prod_{\gamma}^{m/4}\frac{1}{\lambda_\gamma^2(\lambda_\gamma^2+1/4)}\frac{\text{det}[G^+_{XXX}]}{\text{det}[G^-_{XXX}]}\\\nonumber
&=&\sum_{m}^{L/2}\frac{\alpha^{2m}}{\left(1+2\alpha^2\right)^{L/2}}\sum_{\text{states}}\prod_{j=1}^{n}\sin^2{(k_j)}\prod_{i=1}^{n^\pi}\cos^2{(k_l)}\\\nonumber
&=&\sum_{m}^{L/2}\frac{\alpha^{2m}}{\left(1+2\alpha^2\right)^{L/2}}\sum'_{\text{states}}\prod_{j=1}^{n}\left[\sin^2{(k_j)}+\cos^2{(k_l)}\right]\notag\\
&=&\sum_{m}^{L/2}\frac{\alpha^{2m}}{\left(1+2\alpha^2\right)^{L/2}}\binom{L/2}{m}=1.\label{Completeness}
\end{eqnarray}
In going to the third line we have used the completeness of the states for the dimer $XXX$ overlap and the fact that the spin and charge systems are decoupled. After this we take $k_j\in (-\pi,-\pi/2]\cup [0,\pi/2]$ so that we can independently choose it to represent either a $\{k_j,-k_j\}$ pair or a $\{k_j,\pi-k_j\}$ pair. This  allows us to rearrange the terms of the sum to be over the possible choices for the $m=N/2$ $k$'s within this fundamental region,  denoted by $\sum'_\text{states}$ and weighted by their combined overlap, $\sin^2{(k_j)}+\cos^2{(k_l)}=1$ .



\subsection{Finite $U$ Overlap}

Having derived the overlap in the simpler case $U=\infty$ we now move on and consider its calculation for finite $U$. To do this end we follow the approach of Gombor and Bajnok~\citep{GB1, GB2}. As we shall see, this approach does not provide an exact derivation of the ovelaps but rather a well founded conjecture on their form. The idea is as follows: since we are dealing with integrable initial states we assume that, because of their special nature, the overlap formula will factorize into the following form (cf.~\eqref{InftyOverlap})
\begin{eqnarray}\label{GeneralOverlap}
\!\!\!\!\!\!\frac{|\braket{\Psi_0}{\{k_j,-k_j\}\{\lambda_\gamma,-\lambda_\gamma\};m}|^2}{\|\ket{\{k_j,-k_j\}\{\lambda_\gamma,-\lambda_\gamma\};m}\|^2}=\mathcal{A}_m\prod_{j=1}^{N/2}\mathfrak{h}_c(k_j)\prod_{\gamma}^{M/2}\mathfrak{h}_s(\lambda_\gamma)\frac{\text{det}[G^+]}{\text{det}[G^-]},
\end{eqnarray}
where $\mathfrak{h}_{c,s}(x)$ are as yet unknown single particle overlap functions, $G^\pm$ are Gaudin matrices similar to~\eqref{InftyGaudinMatrix} and $\mathcal{A}_m$ is some combinatorial factor coming from the action of lowering operators, e.g. $\eta^-$ or $S^-$ acting on the initial state.  At this point two remarks are in order: 
\begin{itemize}
\item[(i)]The form \eqref{GeneralOverlap} has been shown to arise in all cases were exact overlaps between Bethe states and integrable states can been calculated exactly~\cite{Pozsgay2014, BrockmannDenardisWoutersCaux, Brockmann2, BrockmannDenardisWoutersCaux2, JiangPozsgay}. In fact, comparisons with exact numerics in small systems showed that it holds for essentially all integrable states in integrable models~\cite{deLeeuw, de2015one, buhl2016one, foda2016overlaps, pozsgay2018overlaps, GB1, GB2}.  
\item[(ii)] In writing \eqref{GeneralOverlap} we assumed that only states with pairs of opposite rapidities have non-zero overlap with an integrable state. In essence, this constraint comes from the request that the integrable state is annihilated by all conserved charges that are odd under spatial reflection~\cite{PiroliPozsgayVernier}.  Even though the opposite rapidity pairing is the generic way to fulfil this reflection symmetry condition~\cite{PiroliPozsgayVernier}, it is not the only one~\cite{FrolovQuinn,Beisert}. For instance, in the previous section we saw that in the strong coupling limit of Hubbard we also have pairs of the form $\{k,\pi-k\}$ (alternative pairings for the $U=\infty$ Hubbard have been observed also for different initial states~\cite{BertiniTartagliaCalabrese, TartagliaCalabreseBertini}). Here we note that for Hubbard such non-generic parings are a feature of the strong-coupling limit and do not correspond to consistent solutions of the Bethe equations for finite $U$. This can be explicitly checked by inserting such pairs into~\eqref{BAE1} and \eqref{BAE2} for $N=2,4,\dots $. Upon doing so it is seen that only the allowed pair of this form is $\{0,\pi\}$ if it is also accompanied by $\lambda=0$.
\end{itemize}

A crucial feature of Eq.~\eqref{GeneralOverlap} is that, since in the thermodynamic limit the ratio of the Gaudin determinants approaches $1$, the extensive part of the overlap is given by the functions $\mathfrak{h}_{c,s}(x)$. Thus, it can be determined by a low-density calculation as we now explicitly show. 

We begin by noting that our initial state is a spin singlet and so a nonzero overlap between it and an $N$ particle state requires $M=N/2$ which we take to be even (the $M$ odd case can be treated in the same fashion). Directly computing the overlap for the highest weight state we find
\begin{eqnarray}\nonumber
\braket{\Psi_0}{\{k_j,-k_j\}\{\lambda_\gamma,-\lambda_\gamma\}}&=&\frac{\alpha^{N/2}}{(1+4\alpha^2)^{L/4}}\sum^{L/2}_{l_1< l_2\dots< l_{N/2}}\sum^L_{n_1\leq n_2\dots\leq n_N}\sum_{P\in \mathcal{S}_N}(-1)^Pe^{i\sum_j k_{P_j}n_j}
\\
&&\times\prod_{j=1}^{N/2}\left[\delta_{n_{2j-1},2l_j-1}\delta_{n_{2j},2l_{j}}+\frac{1}{2}\left(\delta_{2l_j-1,n_{2j-1}}\delta_{2l_j-1,n_{2j}}-\delta_{2l_{j},n_{2j-1}}\delta_{2l_{j},n_{2j}}\right)\right]
\varphi^\text{d}_P\,,
\end{eqnarray}
where the product over Kronecker deltas comes from the charge part of the overlap and 
\begin{eqnarray}
\varphi^\text{d}_P\equiv \sum_{\vec{a}}\prod_{j=1}^{N/2}\epsilon^{2j-1, 2j}\varphi_P(\vec{a}).
\end{eqnarray}
This latter term is the overlap between a dimer state and the eigenstates of an inhomogeneous $XXX$ spin chain with the arrangement of inhomogeneities depending on the permutation $P$. In the $U\to\infty$ limit the inhomogeneities vanish and it reduces to the previous result of $\varphi_{XXX}^\text{d}$. Using the form of the wavefunction we can write down this factor explicitly 
\begin{eqnarray}
\varphi^\text{d}_P=\sum_{Q\in S_M}\prod_{1\leq Q_\gamma <Q_\beta\leq M}\frac{\lambda_{Q_\gamma}-\lambda_{Q_\beta}-i U}{\lambda_{Q_\gamma}-\lambda_{Q_\beta}}\prod^{N/2}_{l=1}\frac{s_{P_{2l-1}}-s_{P_{2l}}+i U}{\lambda_{Q_l}-s_{P_{2l}}+iU/2}F_P(\lambda_{Q_l},2l-1).
\end{eqnarray}
From this expression we note that given two permutations $P,~P'$ which differ by exchanging adjacent spins $P_{2j-1}\leftrightarrow P_{2j}$ the overlaps are related by a phase factor 
\begin{eqnarray}
\varphi^\text{d}_{P'}&=&-\frac{s_{P_{2l-1}}-s_{P_{2l}}-i U}{s_{P_{2l-1}}-s_{P_{2l}}+i U}\varphi^\text{d}_{P} = -e^{2i\Phi(k_{P_{2l-1}},k_{P_{2l}})}\varphi^\text{d}_{P},
\end{eqnarray}
which is the bare scattering phase shift between two particles of opposite spin.  We proceed as we did in the $U\to \infty $ case by splitting the sum over permutations into one over different classes which contain a certain permutation $P$ and all others which can be obtained from it by exchanges of the form $P_{2j-1}\leftrightarrow P_{2j}$.  
Our overlap can then be expressed as a sum over these classes
\begin{eqnarray}
\label{exact}
\braket{\Psi_0}{\{k_j,-k_j\}\{\lambda_\gamma,-\lambda_\gamma\}}&=&\frac{\alpha^N}{(1+4\alpha^2)^{L/2}}\sum_{P\in \mathcal{S}_N}(-1)^P\Gamma_P(\vec{k}) \Delta_P(\vec{k}) \varphi^\text{d}_P\,,
\end{eqnarray} 
where we introduced
\begin{eqnarray}
\Delta_P(\vec{k})&=&\prod_{l=1}^{N/2}\left[\frac{1}{2}\left(e^{-ik_{P_{2l-1}}}+e^{-ik_{P_{2l}}+2i\Phi(k_{P_{2l-1}},k_{P_{2l}})}\right)\right.\notag\\
&&\qquad\qquad\left.+\left(\frac{1-e^{i (k_{P_{2l}}+ k_{P_{2l-1}})}}{4}\right)\left(1+e^{2i\Phi(k_{P_{2l-1}},k_{P_{2l}})}\right)\right],\\
\Gamma_P(\vec{k})&=&\sum^{L/2}_{l_1<l_2\dots <l_{N/2}}e^{2i\sum_j^{N/2}(k_{P_{2j-1}}+k_{P_{2j}})l_j}.
\end{eqnarray}
This form is similar to the one we found for $U=\infty$ with the significant difference that the spin part now depends on the permutation preventing us from summing the terms as we did before. The function $\Gamma_P(\vec{k})$ will be largest if the phase in the summand is just $1$, i.e., if the momenta are paired as $k_{P_{2j-1}}=-k_{P_{2j}}$.  Other terms are $\mathcal{O}(1/L^2)$ with respect to these. Moreover, for such a paring of homogeneities the spin factor has been calculated~\cite{GP}
\begin{eqnarray}
\varphi^\text{d}_P=\prod_{\nu=1}^{N/4}\frac{U^2}{\lambda_\nu\sqrt{\lambda_\nu^2+(U/2)^2}}\prod_{1\leq\gamma<\beta\leq N/2}\frac{(\lambda_\gamma-\lambda_\beta)^2+U^2}{(\lambda_\gamma-\lambda_\beta)^2}\text{det}[G^{+}_{iXXX}],
\end{eqnarray} 
where $G^{+}_{iXXX}$ is the Gaudin matrix for the inhomogeneous chain.  From the above expression we see that the dependence on the inhomogeneities appears only through the Gaudin matrix and moreover this is independent of any particular ordering of the inhomogeneities so that $\varphi^\text{d}_P=\varphi^\text{d}_{P'}$. This then allows us to use the sum over permutations to rewrite the ordered sum over $n_j$ into an unordered one as we did  before.
Using this we  get
\begin{eqnarray}\nonumber
\braket{\Psi_0}{\{k_j,-k_j\}\{\lambda_\gamma,-\lambda_\gamma\}}&= &\frac{\alpha^{N/2}}{(1+4\alpha^2)^{L/4}}L^N\prod_{l=1}^{N/2}e^{i\Phi_{k_l}}\left[ \cos{(k_l+\Phi_{k_l})}
\right]\varphi^\text{d}_{P}\\
&&+\dots
\end{eqnarray}
where we have introduced the	phase shift $\Phi_k=\frac{\pi}{2}-\arctan{\left(\frac{2s_k}{U}\right)}$. The ellipsis refers to terms coming from the permutations which do not have the pairing structure of the homogeneities and which are lower order in $L$. At low density the norm of the Bethe state has the leading term
\begin{eqnarray}
\!\!\!\!\|\ket{{\{k_j,-k_j\}\{\lambda_\gamma,-\lambda_\gamma\}}}\|^2=U^{N/2}L^N\!\!\!\prod_{1\leq\gamma<\beta \leq N/2}\left[\frac{(\lambda_\gamma-\lambda_\beta)^2+U^2}{(\lambda_\gamma-\lambda_\beta)^2}\right]\text{det}[G_{iXXX}],
\end{eqnarray}
where we neglected sub-leading contributions in $L$. Combining these two formulae together we have that the normalized squared overlap in the low density limit $ N\ll L$ is 
\begin{eqnarray}\nonumber
\frac{|\braket{\Psi_0}{\{k_j\}\{\lambda_\gamma\}}|^2}{\braket{\{k_j\}\{\lambda_\gamma\}}{\{k_j\}\{\lambda_\gamma\}}}&\approx&\frac{\alpha^N}{(1+4\alpha^2)^{L/2}}\prod_{l=1}^{N/2}| \cos{(k_l+\Phi_{k_l})}
|^2\prod_{\alpha}^{N/4}\frac{U^4}{\lambda_\gamma^2{(\lambda_\gamma^2+(U/2)^2)}}.
\end{eqnarray}
From this we can read off the single particle overlap functions and overall constant factor
\begin{eqnarray}
\mathcal{A}_0 = \frac{\alpha^N}{(1+4\alpha^2)^{L/2}},\quad
\mathfrak{h}_c(k)=|\cos{(k+\Phi_{k})}|^2,\quad 
\mathfrak{h}_s(\lambda)=\frac{U^4}{\lambda^2{(\lambda^2+(U/2)^2)}}.
\end{eqnarray}
Plugging now in \eqref{GeneralOverlap} we find the following conjecture for the overlap formula at arbitrary density and interactions strength
\begin{eqnarray}\label{Overlap}
\!\!\!\!\!\!\!\!\!\frac{|\braket{\Psi_0}{\{k_j\}\{\lambda_\gamma\}}|^2}{\|\ket{\{k_j\}\{\lambda_\gamma\}}\|^2}&=&\frac{\alpha^N}{(1+4\alpha^2)^{L/2}}\prod_{l=1}^{N/2}\frac{s_{k_l}^2[c_{k_l}+U/2]^2}{s_{k_l}^2+(U/2)^2}\prod_{\gamma}^{N/4}\frac{U^4}{\lambda_\gamma^2{(\lambda_\gamma^2+(U/2)^2)}}\frac{\text{det}[G^+]}{\text{det}[G^-]},
\end{eqnarray}
where we used the shorthand notation \eqref{eq:shorthand} and introduced the $\frac{3}{4}N\times \frac{3}{4}N$ Gaudin matrices given by (cf.~\eqref{GaudinMatrix} and \eqref{InftyGaudinMatrix})
\begin{align} \label{UGaudinMatrix}
G^\pm_{i,j}&=\delta_{ij}\left[\frac{L}{\cos{(k_j)}}+\sum_{\gamma=1}^{N/4}\phi^\pm_1(\lambda_\gamma,s_j)\right], \\
G^\pm_{i,N/2+\gamma}&= G_{N/2+\gamma,i}=-\phi^\pm_1(\lambda_\gamma-s_j), \\
G^\pm_{N/2+\gamma,N/2+\beta}&=\delta_{\gamma \beta}\left[\sum_{j=1}^{N/2}\phi^\pm_1(\lambda_\gamma,s_j)-\sum_{\gamma=1}^{N/4}\phi^\pm_2(\lambda_\gamma,\lambda_\gamma)\right]+\phi^\pm_2(\lambda_\gamma,\lambda_\beta),
\end{align}
with $i,j\in[1,N/2]$, $\gamma,\beta\in[1,N/4]$ and $\phi^\pm_n(\lambda,\mu)=\phi_n(\lambda-\mu)\pm \phi_n(\lambda+\mu)$. This expression is valid for momenta and spin rapidities paired as $\{k_j,-k_j\}$ and $\{\lambda_\gamma,-\lambda_\gamma\}$. The $U=\infty$ result of the previous subsection can be recovered from this formula after an appropriate scaling of the rapidities. 
 
 The overlap with the descendant states $\ket{\{k_j\}\{\lambda_\gamma\};m}$ can be calculated in a similar fashion.  For instance, consider $\matrixel{\Psi_0}{\eta^-}{\{k_j\}\{\lambda_\gamma\}}$ and act with the lowering operator to the left on the initial state instead of the eigenstate. Using $\bra{K_l}\eta^-=(-2\alpha) \bra{0}_{2l-1}\otimes \bra{0}_{2l}$ we have that the resulting state takes the same form as $\bra{\Psi_0}$ but with modified coefficients whose overlap can then be calculated as detailed above. For generic $m \leq (L-N)/2$ this procedure yields 
\begin{eqnarray}
\label{eq:overlapfull}
\!\!\!\!\frac{|\braket{\Psi_0}{\{k_j\}\{\lambda_\gamma\};m}|^2}{\|\ket{\{k_j\}\{\lambda_\gamma\};m}\|^2}\!\!&=&\!\!\frac{\mathcal{A}_m\alpha^N}{(1+4\alpha^2)^{L/2}}\prod_{l=1}^{N/2}\frac{s_{k_l}^2[c_{k_l}+U/2]^2}{s_{k_l}^2+(U/2)^2}\prod_{\gamma=1}^{N/4}\frac{U^4}{\lambda_\gamma^2{(\lambda_\gamma^2+(U/2)^2)}}\frac{\text{det}[G^+]}{\text{det}[G^-]},
\end{eqnarray}
with 
\begin{eqnarray}
\label{Combinatorial}
\mathcal{A}_m&=&(2\alpha)^{2m}\frac{(L-N-m)!}{m!(L-N)!}\left[\frac{\left(\frac{L-N}{2}\right)!}{\left(\frac{L-N}{2}-m\right)!}\right]^2.
\end{eqnarray}
Note that only the overlap with the state with $N+2m=L$ survives in the $\alpha\to\infty$ limit.

\section{Quench Action}\label{SectionV}
Having obtained an expression for the overlap of the Bethe states and the initial state we will now turn to the study of the quench dynamics using the Quench Action formalism~\cite{CauxEssler}. The technique is reviewed pedagogically elsewhere~\cite{Caux} but, for completeness, we give a brief description here.  At its heart, the method relies upon the existence of a \emph{representative eigenstate} of the post-quench Hamiltonian that captures the evolution of local operators in the thermodynamic limit. More precisely, denoting by $\mathcal{O}(x)$  a generic local observable, the representative eigenstate $|\Phi{{\rangle}_{L}}$ of $H$ is chosen in such a way that
\begin{equation}
\label{Eq:REAtimedep}
\lim_{\rm th}\frac{\tensor*[_{L}]{{\braket*{\tilde \Psi_0}{\mathcal{O}(x,t)|\tilde \Psi_0}}}{_{L}}}{\tensor*[_{L}]{\braket*{\tilde \Psi_0}{\Psi_0}}{_{L}}}=\lim_{\rm th}\frac{1}{2}\left[\frac{\tensor*[_{L}]{\braket*{\tilde \Psi_0}{\mathcal{O}(x,t)|\Phi}}{_{L}}}{\tensor*[_{L}]{\braket*{\tilde\Psi_0}{\Phi}}{_{L}}}+\frac{\tensor*[_{L}]{\braket*{\Phi}{\mathcal{O}(x,t)|\tilde \Psi_0}}{_{L}}}{\tensor*[_{L}]{\braket*{\Phi}{\tilde \Psi_0}}{_{L}}}\right],
\end{equation}
from which one finds~\cite{CauxEssler, Caux} 
\begin{equation}
\lim_{t\to\infty} \lim_{\rm th}\frac{\tensor*[_{L}]{\braket*{\tilde \Psi_0}{\mathcal{O}(x,t)|\tilde \Psi_0}}{_{L}}}{\tensor*[_{L}]{\braket*{\tilde \Psi_0}{\tilde \Psi_0}}{_{L}}} = \lim_{\rm th}\frac{\tensor*[_{L}]{\braket{\Phi}{\mathcal{O}(x,0)|\Phi}}{_{L}}}{\tensor*[_{L}]{\braket{\Phi}{\Phi}}{_{L}}}.
\label{eq:statval}
\end{equation}
From this equation we see that $|\Phi{{\rangle}_{L}}$ describes the stationary values of all local observables.

In the thermodynamic limit, one can obtain the root densities characterizing the representative state by finding the eigenstate of $H$ that gives the dominant contribution its norm 
\begin{equation}
\braket*{\tilde\Psi_0}{\tilde \Psi_0}= \sum_{\text{states}} |\braket*{n}{\tilde \Psi_0}|^2\,.
\end{equation}
Exchanging the sum over states for a functional integral over the real rapidity distributions we have 
\begin{eqnarray}
\braket*{\tilde\Psi_0}{\tilde \Psi_0} \!\to\! \int \!\!D\rho(k) \!\prod_{n=1}^\infty \!D\sigma_n(\lambda)D\sigma'_n(\lambda)e^{LS[\rho(k),\sigma_n(\lambda),\sigma'_n(\lambda)]}  |\!\braket*{\rho(k),\sigma_n(\lambda),\sigma'_n(\lambda)}{\tilde \Psi_0}|^2
\end{eqnarray}
where the functional  $S[\rho(k),\sigma_n(\lambda),\sigma'_n(\lambda)]$ counts the number microstates with non-zero overlap with $\ket{\Psi_0}$ corresponding to a particular set of distributions. In the thermodynamic limit the functional integral is evaluated in the saddle point approximation. Namely one has to solve 
\begin{eqnarray}
\frac{\delta
S^{QA}}{\delta \rho}\Big |_{\rho,\sigma_n,{\sigma'}_n=\rho^*,\sigma^*_n,{\sigma'}^*_n}=\frac{\delta
S^{QA}}{\delta \sigma_n}\Big |_{\rho,\sigma_n,{\sigma'}_n=\rho^*,\sigma^*_n,{\sigma'}^*_n}=\frac{\delta
S^{QA}}{\delta {\sigma'}_n}\Big |_{\rho,\sigma_n,{\sigma'}_n=\rho^*,\sigma^*_n,{\sigma'}^*_n}=0.
\label{eq:saddlepoint}
\end{eqnarray}
where we introduced the eponymous Quench Action 
\begin{eqnarray}
\!\!\!\!\!S^{QA}[\rho(k),\sigma_n(\lambda),\sigma'_n(\lambda)]=-\frac{2}{L}\mathfrak{R}\left[\log\braket*{\tilde \Psi_0}{\rho(k),\sigma_n(\lambda),\sigma_n(\lambda)}\right]\!-\!S[\rho(k),\sigma_n(\lambda),\sigma'_n(\lambda)].
\end{eqnarray}
In the following subsections we use \eqref{eq:saddlepoint} to determine the rapidity distributions of the representative eigenstate reached after quenches from the states \eqref{eq:ourinitialstate} to the Hubbard model for both infinite and finite $U$.


\subsection{Infinite $U$ steady state}\label{SectionVI}

Let us first deal with the $U\to\infty$ limit, which, as we shall see, represents a somewhat singular point of our quench in a number of respects. As we have seen in Sec.~\ref{sec:infiniteUover}, in this limit the initial state allows for an unusual pairing structure in the momenta. That is, while the spin rapidities appear in pairs of opposite sign $\{\lambda_\gamma,-\lambda_\gamma\}$ the momenta are either paired as $\{k_j,-k_j\}$ which is standard or with a shift of $\pi$, $\{k_j,\pi-k_j\}$. This is related to the existence of a number of additional conserved charges which exist only in this limit~\cite{fagotti2014on, bertini2015pre} and requires some slight modifications  of the Quench Action method. These subtleties are absent at finite $U$ but nevertheless it is instructive to fully understand them before moving on. 

We shall take advantage of the decoupling of the charge and spin degrees of freedom and note that the quench dynamics of the latter are exactly those of the dimer to $XXX$ chain which has been studied before. The dynamics of the charge degrees of freedom, however, can be understood by using the following Hamiltonian describing a single species of spinless fermions
\begin{eqnarray}
\label{eq:Hfree}
H'=\sum_{j=1}^L-\mathfrak{t}\left( c^\dag_{j }c^{\phantom{\dag}}_{j+1 }+c^\dag_{j+1 }c^{\phantom{\dag}}_{j }\right),
\end{eqnarray}
and quenched from the initial state 
\be
\ket{\Psi'_0}=\frac{1}{(1+\alpha^2)^{L/4}}\exp\left[\alpha\sum_{j}^{L/2}c^\dag_{2j-1}c^\dag_{2j}\right],
\ee
which is \emph{not} Gaussian for the fermions $\{c^\dag_{j },c_{j }\}$. 

The overlaps between the eigenstates of $H'$ and $\ket{\Psi'_0}$ have the same form as \eqref{inftyoverlap} but without the spin part and so we can use this simplified setting to understand the dynamics on our system. We can straightforwardly calculate any fermion correlation functions in this initial state, in particular after going to Fourier space we find the following conserved two-point functions
\begin{eqnarray}\label{momentumDist}
\matrixel{\Psi_0'}{\tilde{c}^\dag_k \tilde{c}^{\phantom{\dag}}_k}{\Psi_0'}&=&\frac{\alpha^2}{1+\alpha^2},\\
\matrixel{\Psi_0'}{\tilde{c}^\dag_k \tilde{c}^\dag_{\pi-k}}{\Psi_0'}&=&\frac{\alpha}{1+\alpha^2}\cos{(k)}.
\label{eq:nonconserving}
\end{eqnarray}
Moreover, we also find the following conserved four-point function
\begin{eqnarray}\label{fourpoint}
&&\matrixel{\Psi_0'}{\tilde{c}^\dag_k \tilde{c}^\dag_p\tilde{c}^{\phantom{\dag}}_p \tilde{c}^{\phantom{\dag}}_k}{\Psi_0'},
\end{eqnarray}
which has a non-trivial connected part
\begin{eqnarray}\label{fourpointconn}
\frac{L}{2}\matrixel{\Psi_0'}{\tilde{c}^\dag_k \tilde{c}^\dag_p\tilde{c}^{\phantom{\dag}}_p \tilde{c}^{\phantom{\dag}}_k}{\Psi_0'}^{\mathcal C}&=&\frac{\alpha^2}{1+\alpha^2} \left[\frac{1-\cos(k-p)}{2}\right].
\end{eqnarray}
Eq.~\eqref{momentumDist} gives us the distribution of momenta in the steady state, while the fact that the correlators in Eqs.~\eqref{eq:nonconserving} and \eqref{fourpointconn} are nonzero tells us that the momentum distribution is not sufficient to completely characterize the steady state after the quench. This also means that to describe this quench one cannot apply Quench Action in its standard form.

To overcome this issue one has to include a small but finite magnetic field in the Hamiltonian \eqref{eq:Hfree}, so that the two point function \eqref{eq:nonconserving}, which is not invariant under one-site shifts, is not conserved anymore and we can consider the translational invariant state
\be
|{\tilde \Psi'_0}\rangle = \frac{1-T}{\sqrt 2}\ket{\Psi'_0}\,. 
\ee
Even doing this, the Quench Action treatment is non-trivial because of the presence of non-trivial connected four point correlations. To proceed we introduce the  distributions $\rho(k),\rho^\pi(k)$ and $\rho^h(k)$ describing the distributions of momenta $k\in[0,\pi/2]$ which are part of $\{k,-k\}$ pairs ($\rho$), $\{k,\pi-k\}$ pairs ($\rho^\pi$) and the holes. We also introduce the corresponding distributions $\bar{\rho}(k),\bar{\rho}^\pi(k),\bar{\rho}^h(k)$ which have the same meaning but with $k\in[-\pi,-\pi/2]$. These satisfy 
\be
\rho(k)+\rho^\pi(k)+\rho^h(k)=\frac{1}{2\pi},
\ee
and also 
\be
\bar{\rho}(k)+\bar{\rho}^\pi(k)+\bar{\rho}^h(k)=\frac{1}{2\pi},
\ee
that are the noninteracting Bethe equations. Recall that we require this division of the fundamental domain for $k$ so that we can say that each momenta corresponds to either one type of pair or the other or a hole. The Quench Action for this system is then found to be 
\begin{eqnarray}
S^{QA}_\infty&=&-2\int_0^{\frac{\pi}{2}}{\rm d}k\left[\log{\left({\alpha\cos(k)}\right)}\rho^\pi(k)+\log{\left({\alpha\sin(k)}\right)}\rho(k)\right]\\\nonumber
&&-2\int_{-\pi}^{-\frac{\pi}{2}}{\rm d}k\left[\log{\left({\alpha\cos(k)}\right)}\bar{\rho}^\pi(k)+\log{\left({\alpha\sin(k)}\right)}\bar{\rho}(k)\right]\\\nonumber
&&-\int_{0}^{\pi/2}{\rm d}k\left[\frac{1}{2\pi}\log\frac{1}{2\pi}-\rho(k)\log{\rho}(k)-\rho^\pi(k)\log{\rho^\pi(k)}-\rho^h(k)\log\rho^h(k)\right]\\\nonumber
&&-\int_{-\pi}^{-\pi/2}{\rm d}k\left[\frac{1}{2\pi} \log\frac{1}{2\pi} -\bar{\rho}(k)\log{\bar{\rho}(k)}-\bar{\rho(k)}^\pi(k)\log{\bar{\rho}^\pi}(k)-\bar{\rho}^h(k)\log\bar{\rho}^h(k)\right].
\end{eqnarray}
The first and second  lines here come directly form the overlap formula while the other two are the entropy associated to the choices of microstates corresponding the distributions which differs from the standard Yang-Yang entropy. Minimizing this functional we find that the saddle point of the Quench Action and hence the long time steady state has 
\begin{eqnarray}
\zeta(k)&=&\frac{1}{\alpha^2\sin^2{(k)}}=\bar{\zeta}(k),\\
\zeta^\pi(k)&=&\frac{1}{\alpha^2\cos^2{(k)}}=\bar{\zeta}^\pi(k),
\end{eqnarray}
where $\zeta(k)=\rho^h(k)/\rho(k)$, $\zeta^\pi(k)=\rho^h(k)/\rho^\pi(k)$, $\bar \zeta(k)=\bar \rho^h(k)/\bar \rho(k)$, and $\bar \zeta^\pi(k)=\bar \rho^h(k)/\bar\rho^\pi(k)$. Combining these with our noninteracting Bethe equations we find the distributions to be 
\begin{eqnarray}
2\pi \rho(k)=\frac{\alpha^2}{1+\alpha^2}\sin^2{(k)},\quad 2\pi  \rho^\pi(k)=\frac{\alpha^2}{1+\alpha^2}\cos^2{(k)},\quad 2\pi  \rho^h(k)=\frac{1}{1+\alpha^2}.
\end{eqnarray}
Note that these coincide with the connected four point function \eqref{fourpointconn} for $p=-k$ and $p=\pi-k$. Upon summing over both types of particle pairs we can then reproduce the constant mode occupation 
\be
\rho(k)+\rho^\pi(k)=\frac{1}{2\pi}\frac{\alpha^2}{1+\alpha^2}\,,
\ee
obtained above from direct calculation. Alternatively one could directly sum over both types of pairing structures prior to taking the thermodynamic limit as was when showing the completeness of the states~\eqref{Completeness}. This results in a constant $\zeta(k)=1/\alpha^2$ from which we directly reproduce~\eqref{momentumDist} but which cannot provide the  relative occupation of the different pairs.


 \subsection{Finite $U$ Steady state}\label{SectionVII}
In the finite $U$ case we have only one type of pairing structure and so the Quench Action can be applied more straightforwardly. The only complication arises from the fact that we must now consider the descendant states that have overlap with $|{\tilde\Psi_0}\rangle$.  This would require summing over these states also and weighting them by $-\log{\mathcal{A}_m}$ in the Quench Action,  however this combinatorial factor does not admit a pleasant form in the thermodynamic limit.  In particular, using Stirling's approximation we find
\begin{eqnarray}\nonumber
\log{\mathcal{A}_m}&\approx &(L-N-m)\log{(L-N-m)}-(L-N-2m)\log{(L-N-2m)}\\
&&-m\log{(m)}+2m\log{(\alpha)}.
\end{eqnarray} 
 We should then sum over both $m$ and $N$ subject to the restriction  $(N+2m)/L=4\alpha^2/(1+4\alpha^2)$ which in the Quench Action can be imposed by the inclusion of a Lagrange multiplier.  The sums over $m$ and $N$ could then be evaluated by saddle point thereby yielding a nonlinear relationship between the saddle point values $m^*,N^*$ and $\alpha$ and the Lagrange multiplier.  A more straightforward approach which we shall adopt is to absorb these various factors into  a redefinition of $\alpha\to\tilde{\alpha}$ and vary $\tilde{\alpha}$ \textit{post hoc} to fit the desired initial average particle number~\eqref{AverPart}.  Following this logic the Quench Action is now
\begin{eqnarray}\nonumber
\!\!\!\!\!\!\!\!\!\!\!S^{QA}&=&\int^{\pi}_{0} \!\!{\rm d}k\left[g(k)+h(k)\right]\rho(k)\\\nonumber
&&+\sum_{n=1}^\infty\int_{0}^\infty\!\!\! {\rm d}\lambda \Big\{ g_n(\lambda)\sigma_n(\lambda)+ \left[g_n(\lambda)+g'_n(\lambda)+h'_n(\lambda)\right]\sigma'_n(\lambda)\Big\}\\\nonumber
&&-\frac{1}{2}\int_{-\pi}^\pi  {\rm d}k\Big\{(\rho(k)+\rho^h(k))\log(\rho(k)+\rho^h(k))-\rho(k)\log\rho(k)-\rho^h(k)\log\rho^h(k)\Big\}\\\nonumber
&&-\frac{1}{2}\sum_n^\infty\int_{-\infty}^\infty \!\!{\rm d}\lambda \Big\{( \sigma_n(\lambda)+\sigma_n^h(\lambda))\log(\sigma_n(\lambda)+\sigma^h_n(\lambda))-\sigma_n(\lambda)\log\sigma_n(\lambda)-\sigma^h_n(\lambda)\log\sigma^h_n(\lambda)\Big\}\\\label{QuenchActionU}
&&-\frac{1}{2}\sum_n^\infty\int_{-\infty}^\infty \!\!{\rm d}\lambda \Big\{( \sigma'_n(\lambda)+{\sigma'}_n^h(\lambda))\log(\sigma'_n(\lambda)+{\sigma'}^h_n(\lambda))-\sigma'_n(\lambda)\log\sigma'_n(\lambda)-{\sigma'}^h_n(\lambda)\log{\sigma'}^h_n(\lambda)\Big\}.
\end{eqnarray}
In the first line we have the terms coming from the overlap~\eqref{Overlap} and in the remainder we have written out explicitly the Yang-Yang entropy for the Hubbard model.   For convenience we split the overlap functions into the following contributions
\begin{eqnarray}
g(k)&=&\log{\left(\frac{s_k^2+(U/2)^2}{s_k^2}\right)}-\log{(\tilde{\alpha})},\\\nonumber
h(k)&=&-2\log{(c^{\phantom{\dag}}_k+U/2)},\\\nonumber
g_n(\lambda)&=&\sum_{j=1}^n\log\left[\left(\frac{\lambda}{U}+\frac{i}{2}(n+i-2j)\right)^2\left(\left(\frac{\lambda}{U}+\frac{i}{2}(n+i-2j)\right)^2+\frac{1}{4}\right)\right],\\\nonumber
g'_n(\lambda)&=&\sum_{j=1}^{2n}g(k_{j,n}(\lambda)),\\\nonumber
h'_n(\lambda) &=&\sum_{j=1}^{2n} h(k_{j,n}(\lambda)),
\end{eqnarray} 
where we introduced
\be
k_{j,n}(\lambda) = \begin{cases}
\pi-\arcsin{(\lambda+(n+1-j) i U/2)} & j\quad \text{odd}\\
\arcsin{(\lambda+(n-j)i U/2)} & j< 2n\quad \text{even} \\
\pi-\arcsin{(\lambda-n i U/2)} & j= 2n \\
\end{cases}.
\ee
After minimizing $S^{QA}$ we arrive at a set of thermodynamic Bethe ansatz equations which determine $\zeta(k),\eta_n(\lambda)$ and $\eta'_n(\lambda)$. These can then be cast in a partially decoupled form as the Bethe Takahashi equations~\eqref{ContBAE1}-\eqref{ContBAE3}. The result reads as 
\begin{eqnarray}\nonumber
\log[\eta_n(\lambda)]&=&s*\log[1+\eta_{n+1}][1+\eta_{n+1}](\lambda)+d_n(\lambda)-\delta_{n,1}\int_{-\pi}^\pi {\rm d}k\, c_k\, s(\lambda-s_k)\log{[1+\zeta^{-1}(k)]},\\\nonumber
\log[\eta'_n(\lambda)]&=&s*\log[1+\eta'_{n+1}][1+\eta'_{n+1}](\lambda)+d'_n(\lambda)-\delta_{n,1}\int_{-\pi}^\pi {\rm d}k\, c_k\, s(\lambda-s_k)\log{[1+\zeta^{-1}(k)]},\\\label{QATBA}
\log{[\zeta(k)]}&=&d(k)+\int_{-\infty}^\infty {\rm d}\lambda\, s(s_k-\lambda)\log\left[\frac{1+\eta'_1(\lambda)}{1+\eta_1(\lambda)}\right],
\end{eqnarray}
where we used the shorthand notation \eqref{eq:shorthand} and introduced the ratios
\be
\label{eq:etas}
\zeta(k)=\frac{\rho^h(k)}{\rho(k)}, \qquad \eta_n(\lambda)=\frac{{\sigma}^h_n(\lambda)}{\sigma_n(\lambda)},\qquad \eta'_n(\lambda)=\frac{{\sigma'}^h_n(\lambda)}{\sigma'_n(\lambda)}\,.
\ee
The driving terms $d(k),d_n(\lambda)$ and $d'_n(\lambda)$ depend on the explicit form of the overlap equations and are given by
\begin{eqnarray}\nonumber
d(k)&=&g(k)-s*\left[g_1'(s_k)+h'_1(s_k)\right]=-\log{\left[\tanh^2\left(\frac{\pi s_k}{2U}\right)\right]}+h(k)-s*h'_1(s_k)\\\nonumber
d_n(\lambda)&=&g_n(\lambda)-s*\left[g_{n+1}(\lambda)+g_{n-1}(\lambda)\right]=\log{\left[\tanh^2\left(\frac{\pi \lambda}{2U}\right)\right]},\\\nonumber
d'_n(\lambda)&=&g_n(\lambda)-s*\left[g_{n+1}(\lambda)+g_{n-1}(\lambda)\right]+g'_n(\lambda)-s*\left[g'_{n+1}(\lambda)+g'_{n-1}(\lambda)\right]\\
&&+h'_n(\lambda)-s*\left[h'_{n+1}(\lambda)+h'_{n-1}(\lambda)\right],\\\notag
&=&\left[2(-1)^{n+1}+1-\delta_{n,1}\right]\log{\left[\tanh^2\left(\frac{\pi \lambda}{2U}\right)\right]}+\delta_{n,1}\left[h'_1(\lambda)-s*h'_2(\lambda)\right],
\end{eqnarray}
The simplified expressions for these driving terms are found by using the Fourier transforms of the functions $g_n(\lambda)$ and $h_n(\lambda)$. In particular we use the result 
\be
\mathcal{FT}[\log(x^2+a^2)]=-\frac{2\pi}{|\omega|}e^{-|a||\omega|}.
\ee

Note that the driving term for the spin degrees of freedom $d_n(\lambda)$ is the same as for the  dimer to $XXX$ quench~\cite{PiroliPozsgayVernier1} while that of the spin-charge bound states resembles a sum of this with the driving term for the Neel to $XXX$  quench~\cite{Brockmann}. The TBA equations should be supplemented with the appropriate boundary conditions for $\eta_n(\lambda)$ and $\eta_n'(\lambda)$ at $n\to\infty $ which we adopt from those used in analogous dimer or Neel to $XXX$ quenches~\cite{wouters2014quenching, pozsgay2014correlations,mestyan2017exact}, i.e., 
\be
\lim_{n\to\infty}\frac{\eta'_n(\lambda)}{\eta'_{n-2}(\lambda)} = \tilde \alpha^{-2},\qquad\qquad \lim_{n\to\infty}\frac{\eta_n(\lambda)}{\eta_{n-1}(\lambda)} =1\,.
\ee  
The integral equations~\eqref{QATBA} take the same form as TBA equations describing the equilibrium state of the system at finite temperature and can be analyzed using the same methods. In particular, they can be integrated numerically by truncating the system to contain only a finite  number of string types, $N_\text{max}$, imposing a cutoff on the rapidity integrals $\Lambda$ and then proceeding iteratively by switching to Fourier space to compute the convolutions.  To facilitate this it is convenient to make a change of variables for the momenta to a rapidity notation~\cite{FrolovQuinn,IllievskiDenardis} 
\be
k\mapsto z(k) = \begin{cases}
\sin(k)& |k|<\pi/2\\
\sin(\pi-k) & \pi/2\leq |k|<\pi
\end{cases}.
\ee 

\section{Steady state Doublon density}\label{SectionVIII}

\begin{figure*}
 \centering
    (a)%
    \raisebox{-\totalheight+\baselineskip}[0pt][\totalheight]{\includegraphics[width=0.45\columnwidth]{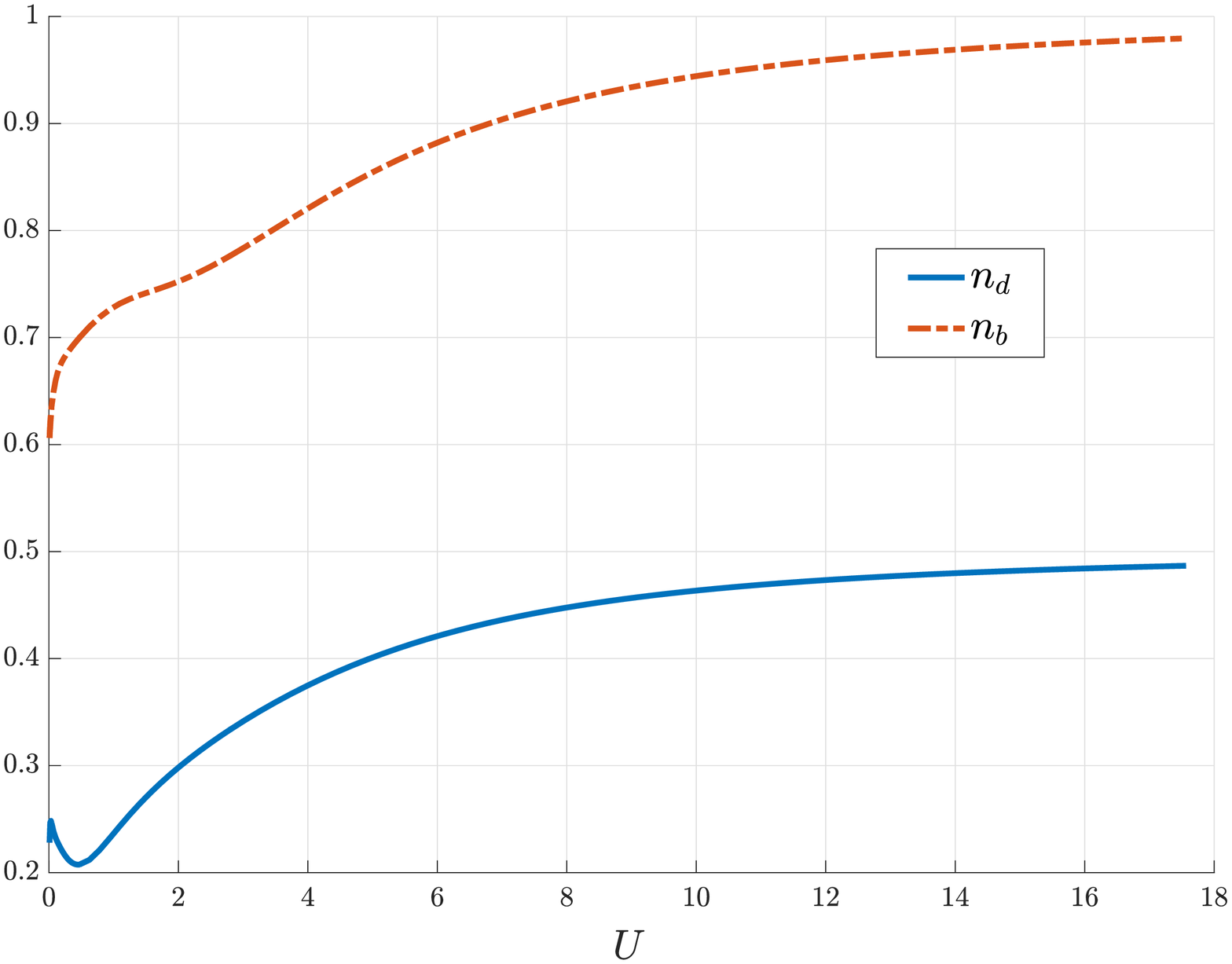}}
    (b)%
    \raisebox{-\totalheight+\baselineskip}[0pt][\totalheight]{\includegraphics[width=0.45\columnwidth]{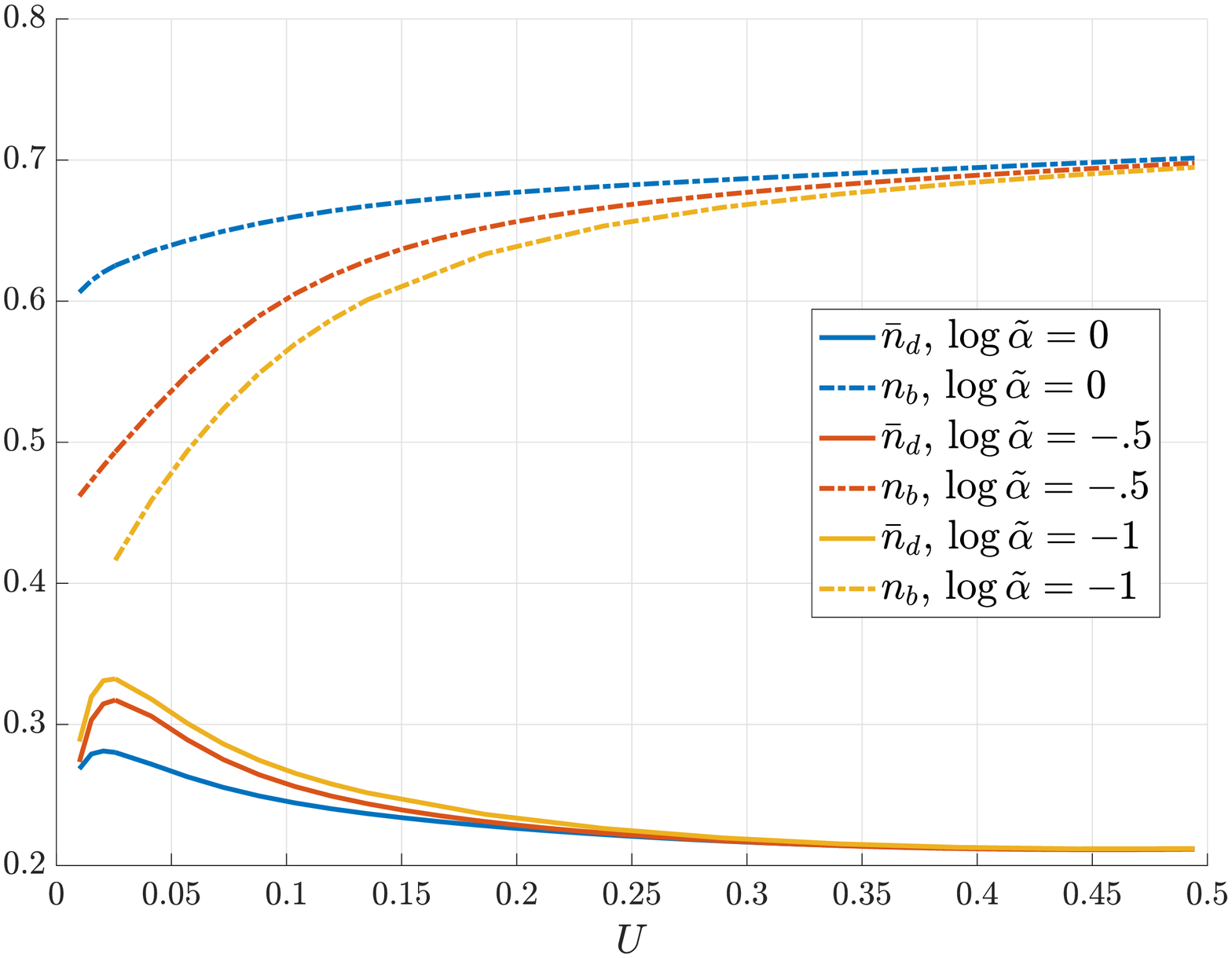}}
\caption{\label{fig:doublon}(a) The density of doublons $n_d$ in the long time steady state as a function of $U$ for $\log{(\tilde{\alpha})}=0$ (solid).  Also shown is the fraction of particles forming bound states  $n_b$ (dashed) in the steady state.  Close to the origin $n_d$ initially increases,  which is hard to discern on this scale.  (b) The normalized doublon density $\bar{n}_d=n_d/n^2$  (solid)  as a function of $U$ close to the origin, up to its minimum  $U\sim .5$ for $\log\tilde{\alpha}=0,-.5,-1$.  Here we see the initial increase.  Also shown is the bound state fraction, $n_b$ (dashed).}
\end{figure*}

In this section we  examine the properties of the steady state for finite $U$ by calculating the expectation value of the number of doubly occupied sites,  known as doublons. This is one of the most readily accessible experimental observables in cold atom experiments~\cite{Doublon1,Doublon2}. The doublon number operator is given by 
\be
\hat{N}_\text{doublon}=\sum_jc^\dag_{j\uparrow}c^{\phantom{\dag}}_{j\uparrow}c^\dag_{j\downarrow}c^{\phantom{\dag}}_{j\downarrow},
\ee
and its expectation value in any eigenstate of the Hubbard model can be evaluated as
\be
 \matrixel{\{k_j\}\{\lambda_\gamma\}}{\hat{N}_\text{doublon}}{\{k_j\}\{\lambda_\gamma\}}= \frac{1}{2\mathfrak{t} }\d{}{U}E(\{k_j\}\{\lambda_\gamma\}) 
\ee
using the Hellmann-Feynman theorem. In particular, employing \eqref{eq:statval} we obtain 
\be
\lim_{t\to\infty} \lim_{\rm th} \frac{\tensor*[_{L}]{\matrixel{\Psi_0}{\hat{N}_\text{doublon}(t)}{\Psi_0}}{_L}}{\tensor*[_{L}]{\braket{\Psi_0}{\Psi_0}}{_L}}=\lim_{\rm th} \frac{\tensor*[_{L}]{\matrixel{\Phi}{\hat{N}_\text{doublon}(t)}{\Phi}}{_L}}{\tensor*[_{L}]{\braket{\Phi}{\Phi}}{_L}}= \lim_{\rm th} \frac{1}{2\mathfrak{t} }\d{}{U}E_{\Phi},
\ee
where $E_{\Phi}$ is the energy of the representative state $\ket{\Phi}_L$. In the thermodynamic limit we can evaluate this derivative as
\begin{eqnarray}
\lim_{\rm th}\frac{1}{2\mathfrak{t} L}\d{}{U}E_{\Phi}
&=&\lim_{\rm th}\frac{1}{L}\sum_{j}\sin(k_j)\d{k_j}{U}+\frac{1}{2\mathfrak{t} L}\sum_{n=1}^\infty\sum_{n=1}^{M'_n}\left[\d{{e_{k-\lambda,n}({\lambda'}_\gamma^n)}}{{\lambda'}^n_\gamma}\d{{\lambda'}_\gamma^n}{U}+\d{{e_{k-\lambda,n}({\lambda'}_\gamma^n)}}{U}\right]\\\label{doublondensity}
&=&\int_{-\pi}^\pi \!\!\!{\rm d}k \sin{(k)}\omega(k)+\sum_{n=1}^\infty\int_{-\infty}^\infty \!\!\frac{{\rm d}\lambda}{2\mathfrak{t} }\left[\d{{e_{k-\lambda,n}({\lambda})}}{\lambda}\mu_n'(\lambda)+\d{{e_{k-\lambda,n}({\lambda})}}{U}\sigma'_n(\lambda)\right]\,\,
\end{eqnarray}
Here we have introduced the distributions $\omega(k), \mu'_n(\lambda)$ which describe the rate of change of the momenta and string centers of the representative state with respect to $U$ in the thermodynamic limit, i.e.  
\be
L^{-1}\sum_j\d{k_j}{U}\to \int {\rm d}k\, \omega(k),
\ee 
and 
\be
L^{-1}\sum_\gamma\d{{\lambda'}^n_\gamma}{U}\to \int {\rm d}\lambda'\, \mu'_n(\lambda')\,.
\ee
Along with the analogous quantity $\mu_n(\lambda)$, $\omega(k)$ and $\mu'_n(\lambda)$ satisfy a set of integral equations obtained by differentiating the logarithm of the Bethe equations~\eqref{BAE1} and \eqref{BAE2}, namely
\begin{eqnarray}\label{BetheHellFeyn1}
&&\!\!\!\!\!\!\!\!\!\![1+\zeta(k)]\omega(k)=\sum_{n=1}^\infty\int_{-\infty}^\infty\frac{{\rm d}\lambda}{2\pi}\phi_n(s_k-\lambda)\left[\mu_n(\lambda)+\mu'_n(\lambda)+\frac{(s_k-\lambda)}{U}\left(\sigma_n(\lambda)+\sigma'
_n(\lambda)\right)\right],\\\label{BetheHellFeyn2}
&&\!\!\!\!\!\!\!\!\!\![1+\eta_n(\lambda)]\mu_n(\lambda)=\!\!\int_{-\pi}^\pi \frac{{\rm d}k}{2\pi} \phi_n(s_k-\lambda)\left[c_k\,\omega(k)-\frac{(s_k-\lambda)}{U}\rho(k)\right]-\sum_{m=1}^\infty T_{nm}*\mu_m(\lambda),\\\label{BetheHellFeyn3}
&&\!\!\!\!\!\!\!\!\!\![1+\eta'_n(\lambda)]\mu'_n(\lambda)=\!\!\int_{-\pi}^\pi \frac{{\rm d}k}{2\pi} \phi_n(s_k-\lambda)\left[\frac{(s_k-\lambda)}{U}\rho(k)- c_k\, \omega(k)\right]-\sum_{m=1}^\infty T_{nm}*\mu'_m(\lambda)-\frac{1}{2\pi}\d{p_{k-\lambda,n}(\lambda)}{U},
\end{eqnarray}
where $\{\rho(k), \sigma_n(\lambda), \sigma'_n(\lambda)\}$ are the saddle point densities and $\{\zeta(k), \eta_n(\lambda), \eta'_n(\lambda)\}$ are the saddle point ratios (cf.~\eqref{eq:etas}) fulfilling~\eqref{QATBA}. 

These Hellman-Feynman equations are of the same form as the Bethe-Takahashi equations presented earlier~\eqref{ContBAE1}--\eqref{ContBAE3} and can also be brought to a partially decoupled form, which is reported in Appendix~\ref{app:decoupledequations}. We integrate these equations numerically using the distributions obtained solving the Quench Action saddle point equations~\eqref{QATBA} and the Bethe-Takahashi equations~\eqref{continuumBAE}. The results are then plugged into~\eqref{doublondensity} to obtain the long time limit of the density of doublons.  We plot the  resulting density   
\be
n_d=\lim_{t\to\infty}\lim_{\rm th} \frac{\tensor*[_{L}]{\matrixel{\Psi_0}{\hat{N}_\text{doublon}(t)}{\Psi_0}}{_L}}{L\tensor*[_{L}]{\braket{\Psi_0}{\Psi_0}}{_L}},
\ee
as a function of $U$  in Figure~\ref{fig:doublon}(a) for $\tilde \alpha=1$ and see that for increasing interaction strength the doublon density approaches $n_d\to 1/2$ i.e.  all particles  forming doublons.  This $U\to \infty$ behaviour is, in fact,  governed by the final term in~\eqref{doublondensity} which in this limit becomes $n_b/2$ where $n_b$ is the fraction of particles which form bound states,
\be
n_b=\frac{1}{N}\sum_{n=1}^\infty 2n \int d\lambda \,\sigma'_n(\lambda).
\ee
This is also plotted in Figure~\ref{fig:doublon}(a) showing the same increase with $U$ as $n_d$ for large $U$.  We understand this by recalling that the bound states have an exponentially decaying wavefunction with localization length $\sim 1/U$ meaning that they become more tightly bound with increasing $U$ resulting in more doubly occupied sites. As the steady state becomes dominated by the bound states at large $U$ the number of doublons therefore also increases.  The other terms in~\eqref{doublondensity} instead govern the small $U$ behaviour with the competition between these two terms resulting in the observed nonmonotonic behaviour. In the limit of $U\to 0$ we can use the decoupling of the spins and the translational invariance of the steady state to determine that $n_d \to n^2/4$ where $n=N/L$ is the particle density.  In  Figure~\ref{fig:doublon} (b) we plot the normalized doublon density close to the origin $\bar{n}_d=n_d/n^2$ for different values of $\tilde{\alpha}$ showing this limiting behaviour.  Also shown is $n_b$ which decreases quickly as $U\to 0$.

\section{Entanglement dynamics}
\label{SectionIX}
\begin{figure}
\centering
\includegraphics[width=0.65\textwidth]{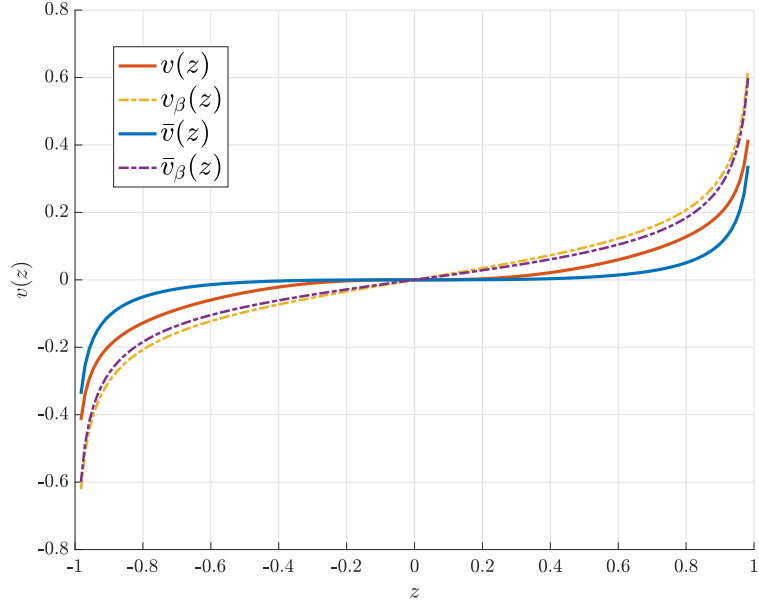}
\caption{The quasiparticle velocities of the charge excitations $v(z), ~\bar{v}(z)$ above the long time steady state for $U=4$ and $\tilde{\alpha},\mathfrak{t}=1$(solid lines). For comparison we also plot the same quantities, denoted $v_\beta(z), ~\bar{v}_\beta(z)$ for the thermal state with inverse temperature $\beta=10$ (dot-dashed lines).   \label{ChargeVelPlots}}
\end{figure}

\begin{figure*}
 \centering
    (a)%
    \raisebox{-\totalheight+\baselineskip}[0pt][\totalheight]{\includegraphics[width=0.45\columnwidth]{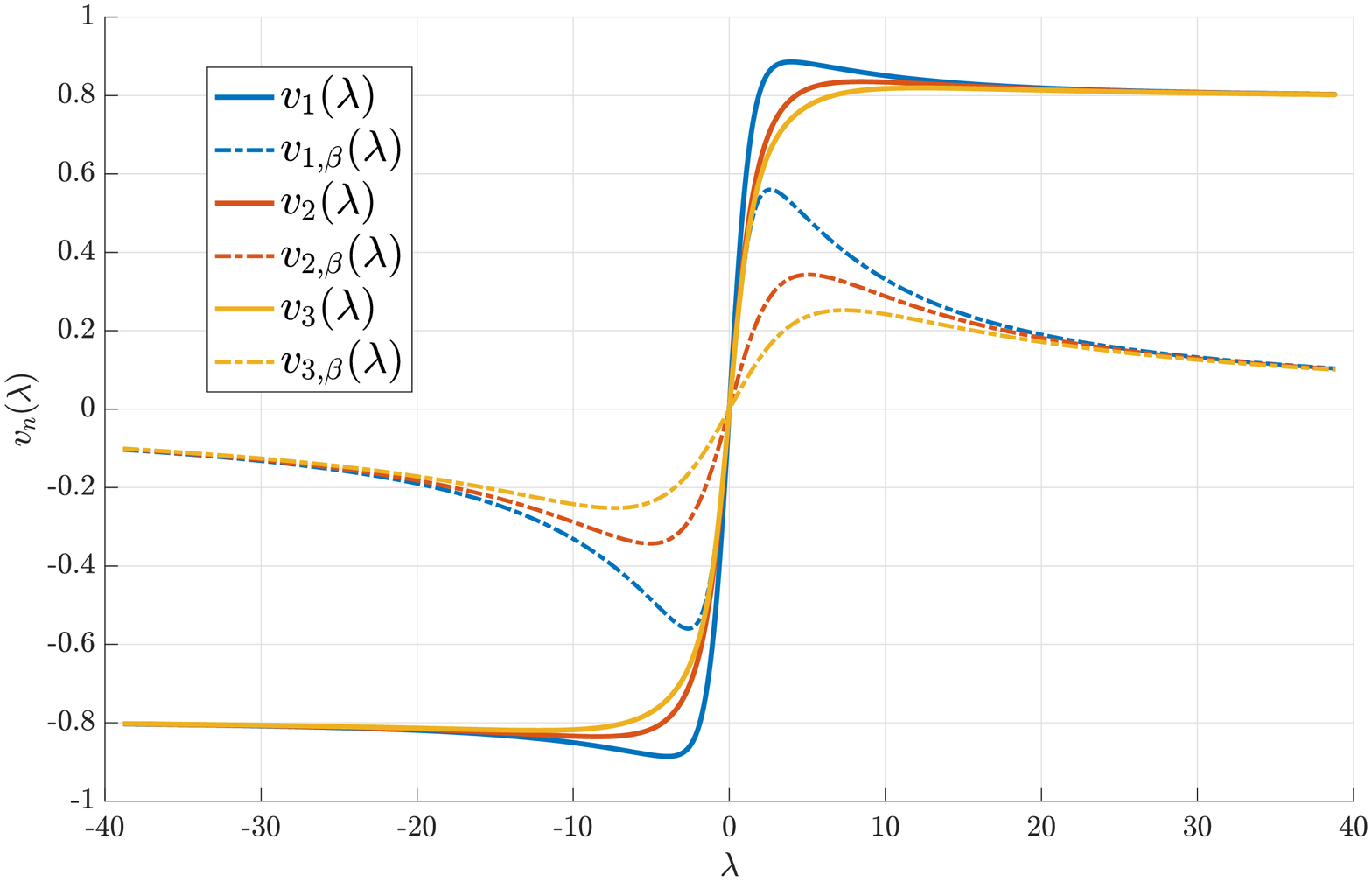}}
    (b)%
    \raisebox{-\totalheight+\baselineskip}[0pt][\totalheight]{\includegraphics[width=0.45\columnwidth]{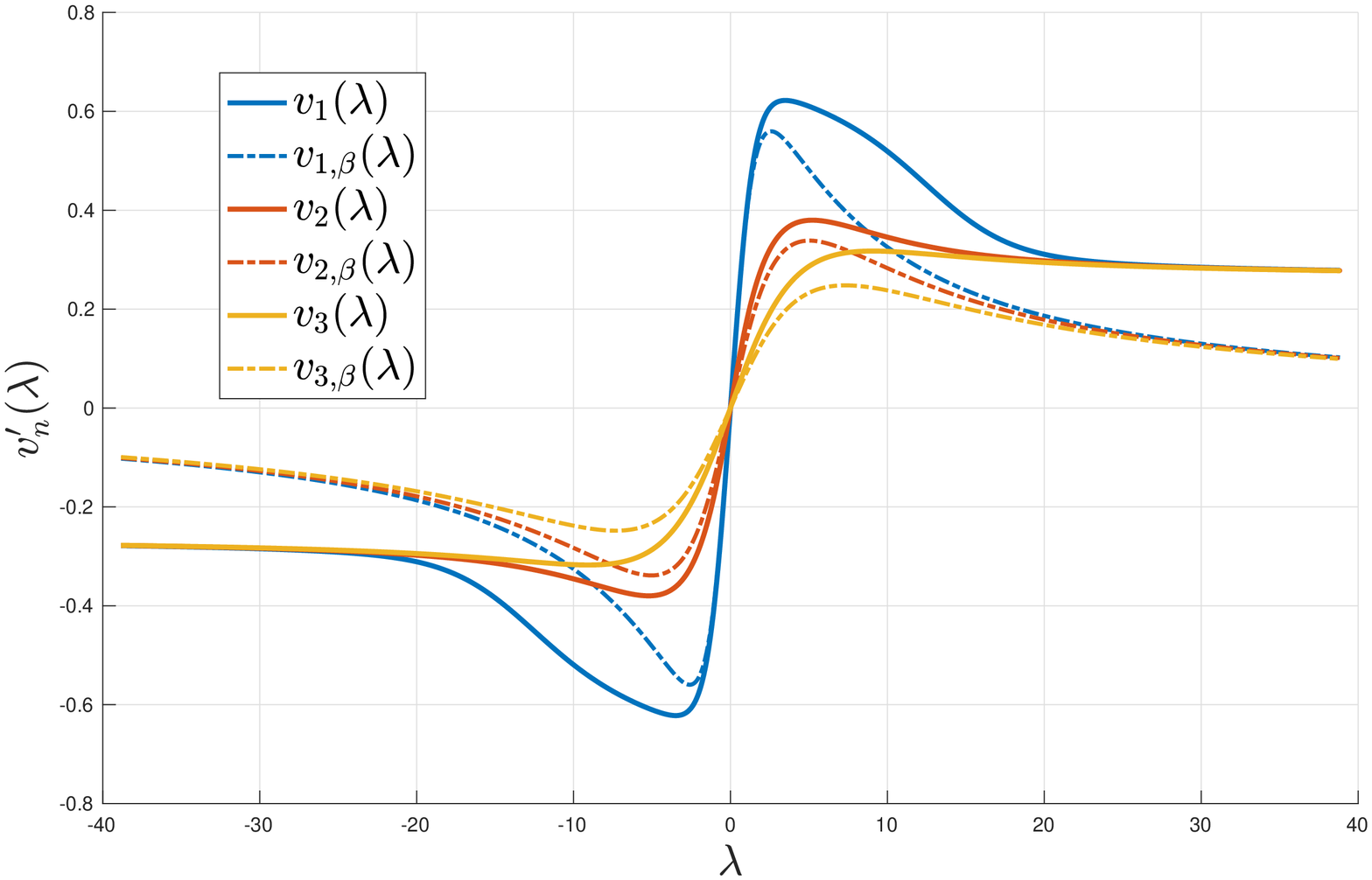}}
\caption{\label{fig:velocities}(a) The first three quasiparticle velocities of the spin excitations $v_n(z), ~n=1,2,3$ above the steady states for $U=4$ and $\tilde{\alpha},\mathfrak{t}=1$ (solid lines).  For comparison we also plot the same quantities, denoted $v_{n,\beta}(\lambda)$ for the thermal state with inverse temperature $\beta=10$ (dot-dashed lines).  (b) The quasi particle velocities of the bound state excitations, $v'_n(z)$, $n=1,2,3$ for the same parameters and compared with the thermal cases denoted $v'_{n,\beta}(\lambda)$.  }
\end{figure*}

A remarkable feature of quenches from integrable initial states is that for these quenches the knowledge of the saddle point state $\ket{\Phi}_L$ gives direct access to the dynamics of bipartite quantum entanglement at the leading order in time~\cite{AlbaCalabrese1, bertini2022growth}. That is, the saddle point state does not only characterise the stationary properties of the system after the quench but also its genuine non-equilibrium dynamics. This striking phenomenon has been first observed in the evolution of the (von Neumann) entanglement entropy~\cite{AlbaCalabrese1} but it has been recently proved to occur for all R\'enyi entropies~\cite{bertini2022growth} and is hence a property of the full entanglement spectrum. In this section we will exploit this feature to characterise the entanglement dynamics from the states~\eqref{eq:ourinitialstate}.

\subsection{Entanglement Entropy}
\label{sec:EE}

The von Neumann entanglement entropy between a subsystem $A$ and its complement is given by 
\begin{eqnarray}
S_{A}(t)=-\text{Tr}_A\left[\varrho_A(t)\log\varrho_A(t)\right]
\end{eqnarray}
where $\varrho_A$ is the reduced density matrix of $A$ and the trace is taken over this region. This is a very difficult quantity to calculate analytically from first principles even for free models.  This problem can be circumvented by a appealing to a phenomenological quasiparticle picture introduced in the context of conformal field theory in~\cite{CalabreseCardy} and later extended to interacting integrable models~\cite{AlbaCalabrese1,AlbaCalabrese2}. Therein the entanglement is viewed as spreading throughout the system by pairs of quasiparticles of  opposite momentum created by the quench which emanate from each point in space.  The region $A$ will become entangled with its complement if it contains one  particle from a pair with the other residing in the complement.  The entanglement entropy is then given summing over all pairs of particles shared between the two regions and weighted by a functional counting their entanglement which on general grounds is argued to be the Yang-Yang entropy in the saddle point state. The resulting expression for the finite time dynamics, valid in the scaling limit $t,\ell\to \infty$ with $t/\ell$ fixed, is 
\begin{figure*}
 \centering
    (a)%
    \raisebox{-\totalheight+\baselineskip}[0pt][\totalheight]{\includegraphics[width=0.45\columnwidth]{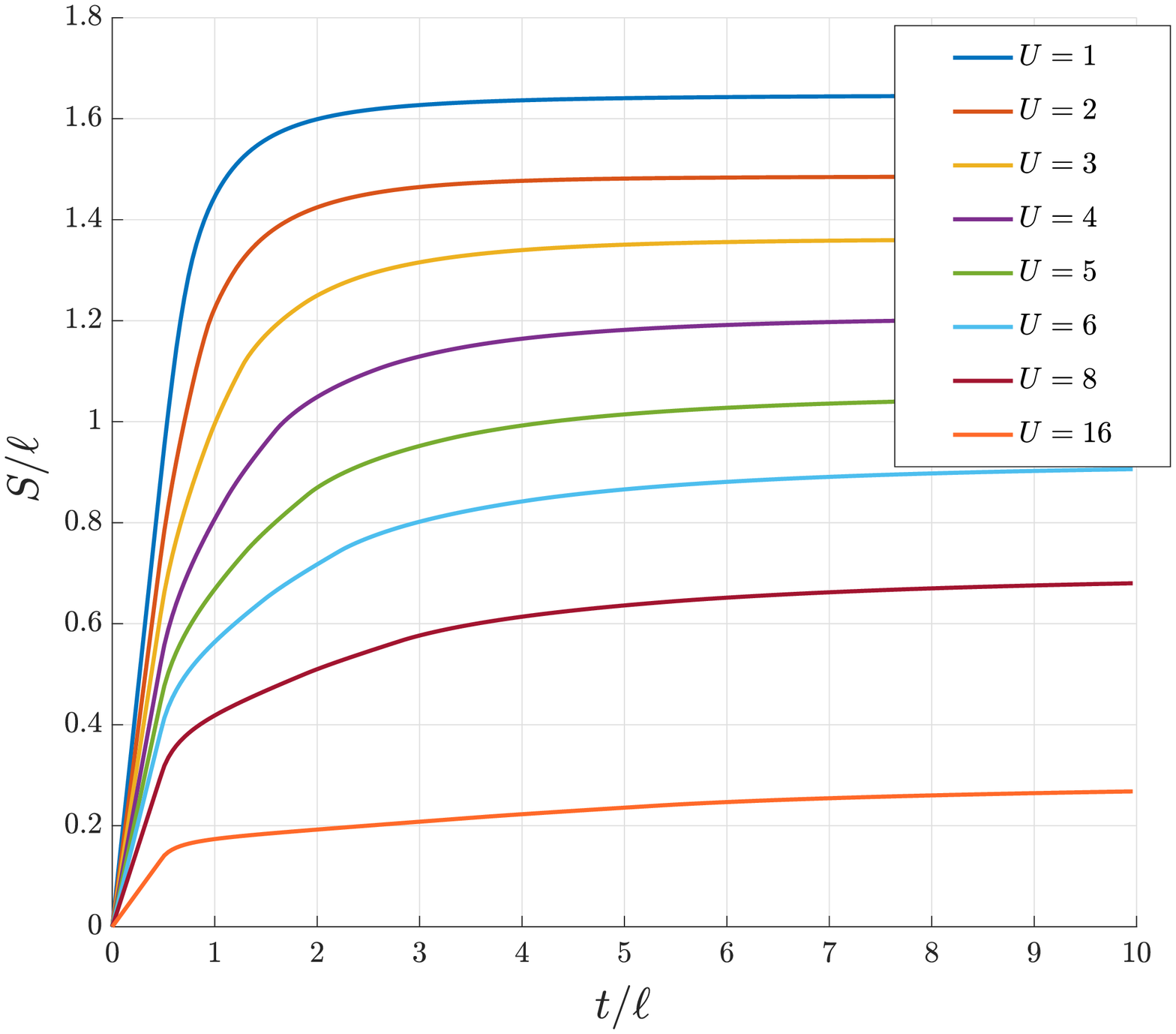}}\hfill
    (b)%
    \raisebox{-\totalheight+\baselineskip}[0pt][\totalheight]{\includegraphics[width=0.45\columnwidth]{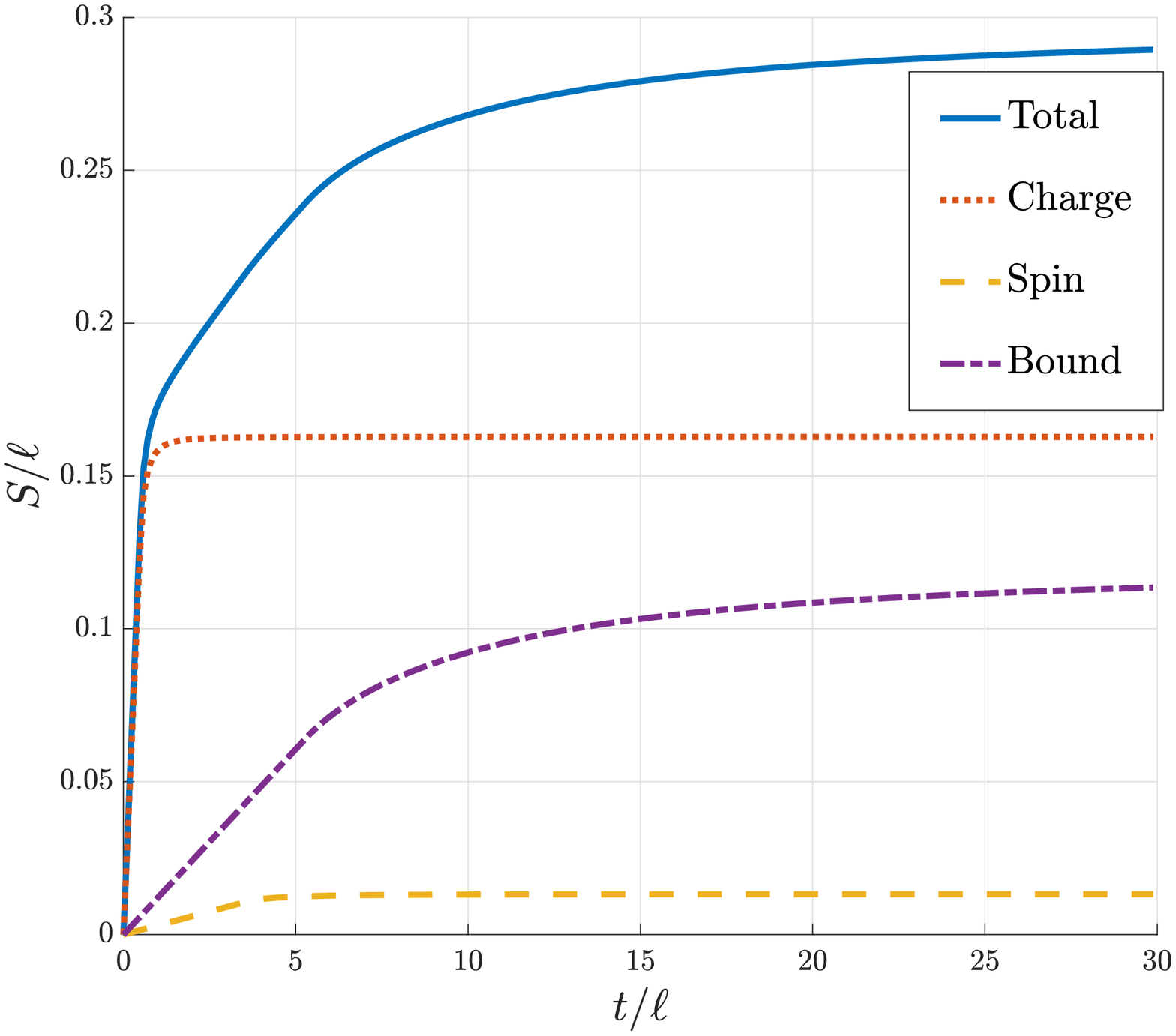}}
\caption{\label{fig:vonNeumann}
(a) The evolution of the entanglement entropy $S/\ell$ as a function of the scaled time $t/\ell$ at different interaction strengths from $U=1$ to $U=16$ with $\tilde{\alpha}=1$. As the interaction increases the slope of the initial linear growth and the height of the plateau decrease.  For intermediate interaction strength a two slope structure develops.  (b) The entanglement entropy  at $U=16,\tilde{\alpha}=1$ (solid line) decomposed into its constituents coming from the charge (dotted line), spin (dashed line) and bound state dynamics (dot-dashed line).  The charge degrees of freedom saturate quickly after which the bound states dominate the entanglement growth
}
\end{figure*}
\begin{eqnarray}\nonumber
S_{A}(t)&=&\int_{0}^\pi {\rm d}k\, s_{\textsc{yy}}(k) \Big\{\ell \theta(2v(k) t - \ell)+2v(k) t \theta(\ell-2t v(k))\Big\}\\\nonumber&&+\sum_{n=1}^\infty\int _{0}^\infty {\rm d}\lambda\, s_{\textsc{yy},n}(\lambda)\Big\{ \ell \theta(2 v_n(k)-\ell)+2 v_n(k) t \theta(\ell-2t v_n(k))\Big\}\\\label{EntEnt}
&&+\sum_{n=1}^\infty\int_{0}^\infty {\rm d}\lambda\, {s}'_{\textsc{yy}, n}(\lambda)\Big\{\ell \theta(2 v'_n(k)-\ell)+2 v_n'(k) t \theta(\ell -2t v'_n(k))\Big\},
\end{eqnarray}
where we considered the case of $A$ being a block of length $\ell$. In \eqref{EntEnt} we introduced the Yang-Yang entropy densities for each string type and the quasi-particle velocities $v(k),v_n(\lambda)$ and $v'_n(\lambda)$. The former are conveniently expressed in terms of the filling functions
\be
\vartheta(k)\equiv \frac{\rho(k)}{\rho(k)+\rho^h(k)},\qquad
\vartheta_n(\lambda) \equiv \frac{\sigma_n(\lambda)}{\sigma_n(\lambda)+\sigma_n^h(\lambda)},\qquad
\vartheta'_n(\lambda) \equiv\frac{\sigma'_n(\lambda)}{\sigma'_n(\lambda)+\sigma_n^{\prime h}(\lambda)},
\label{eq:fillingfunctions}
\ee
as follows
\begin{eqnarray}
s_{\textsc{yy}}(k) &=& -2(\rho(k)+\rho^h(k))\left[ \vartheta(k)\log\vartheta(k)+(1-\vartheta(k))\log(1-\vartheta(k))\right],\nonumber\\
s_{\textsc{yy}, n}(\lambda) &=& -2(\sigma_n(\lambda)+\sigma_n^h(\lambda))\left[ \vartheta_n(\lambda)\log\vartheta_n(\lambda)+(1-\vartheta_n(\lambda))\log(1-\vartheta_n(\lambda))\right]\,,\label{sec:yangyang}\\\nonumber
{s}'_{\textsc{yy}, n}(\lambda) &=& -2(\sigma'_n(\lambda)+\sigma_n^{\prime h}(\lambda))\left[ \vartheta'_n(\lambda)\log\vartheta'_n(\lambda)+(1-\vartheta'_n(\lambda))\log(1-\vartheta'_n(\lambda))\right].
\end{eqnarray}
The latter are defined as the group velocities of excitations created on top of the long time steady state.  As mentioned in Sec.~\ref{sec:TBA}, the bare energy and momentum of an excitation get dressed by the other particles in the system in a way which depends upon the state it is excited above.  Accordingly the quasi particle velocities are also dressed and satisfy the following set of integral equations~\cite{BonnesEsslerLauchli}. In the case of the Hubbard model we have~\cite{QuinnFrolov,   IllievskiDenardis}
\begin{eqnarray}\nonumber
\left[1+\zeta(k)\right]\rho(k)v(k)&=&\frac{\mathfrak{t}\sin{(k)}}{\pi}+\cos(k)\sum_{n=1}^\infty\int \frac{{\rm d}\lambda}{2\pi}\phi_n(s_k-\lambda)\left[\sigma_n(\lambda)v_n(\lambda)+\sigma'_n(\lambda)v'_n(\lambda)\right],\\
\left[1+\eta(\lambda)\right]\sigma_n(\lambda)v_n(\lambda)&=&\int_{-\pi}^\pi \frac{{\rm d}k}{2\pi} \phi_n(s_k-\lambda)\rho(k)v(k)-\sum_{m=1}^\infty T_{nm}*\sigma_m(\lambda)v_n(\lambda),\\\nonumber
\left[1+\eta'_n(\lambda)\right]\sigma'_n(\lambda)v'_n(\lambda)&=&\frac{\mathfrak{t}}{\pi}\mathfrak{R}\left[\frac{\lambda+inU/2}{\sqrt{1-(\lambda+inU/2)^2}}\right]-\int_{-\pi}^\pi \frac{{\rm d}k}{2\pi}\phi_n(s_k-\lambda)\rho(k)v(k)-\sum_{m=1}^\infty T_{nm}*\sigma'_m(\lambda)v_n'(\lambda).\label{DressedV}
\end{eqnarray}
Once again these equations can be recast in the partially decoupled form which are reported in Appendix~\ref{app:decoupledequations}.

Solving these numerically in conjunction with the TBA for the saddle point and the Bethe-Takahashi equations we find the quasi particle velocities for excitations above the steady state.  In Figures~\ref{ChargeVelPlots}, \ref{fig:velocities} (a) and \ref{fig:velocities} (b) we present the quasiparticle velocities for the charge,  spin and bound state excitations at $U=4$, $\tilde{\alpha}=1$ and with the hopping strength $\mathfrak{t}=1$.  For context, we compare them to the same quantities evaluated for a thermal state at the same interaction and hopping strength and inverse temperature $\beta=10$.  From this we see that in the steady state the velocities of the charge excitations are suppressed whereas the velocities of the spin and bound states are enhanced.

Inserting these into~\eqref{EntEnt} we obtain the finite time evolution of the entanglement entropy.  In Figure~\ref{fig:vonNeumann} (a) we plot the resulting dynamics for different values of $U$. For interaction strength less than $U\approx 2$ we see the typical linear growth of the entanglement entropy followed by a plateau whereas for larger values a second linear regime appears prior to the eventual plateau.  As the interaction is further increased the slope of this second period of growth is suppressed and ultimately disappears for $U\to \infty$.  To understand this in Figure~\ref{fig:vonNeumann} (b) we plot $S_A(t)$ at $U=16$ as well as  the contribution of each of its excitation types,  charge, spin and bound state. We see that the initial linear growth is governed by the fastest charge excitations which reach their plateau early. Subsequently the bound states are the dominant contribution to the entanglement growth but their contribution decreases for large interaction strength. Note that spin degrees of freedom contribute the least and the two slope structure cannot be attributed to spin-charge separation which has been seen previously in non-equilibrium scenarios~\cite{MestyanBertiniPiroliCalabrese, ScopaCalabresePiroli}. 

\subsection{R\'enyi entropies}

Alternative measures of the entanglement between a subsystem, $A$ and its compliment are given by the R\'enyi entanglement entropies which are defined as  
\begin{eqnarray}
\label{eq:Renyi}
S_A^{(n)}(t)=\frac{1}{1-n}\log\left[\text{Tr}_A\varrho_A^n(t)\right],
\end{eqnarray}
where $n\in \mathbb{R}$. In the limit $n\to 1$ this recovers the von Neumann entanglement entropy discussed in the previous subsection.  Besides providing full access to the spectrum of $\varrho_A(t)$ --- the entanglement spectrum --- $S_A^{(n\geq2)}(t)$ are of significant interest because they can be measured in actual experiments~\cite{Islam_2015, Kaufman_2016, Linke_2018, Lukin_2019, Elben_2020, Zhou_2020, Neven_2021, kokai2021entanglement,vitale2022symmetry}. Like the von Neumann entropy, $S_A^{(n)}(t)$ is expected to show an initial linear growth with time before eventually saturating.  The long time limit is given by 
\begin{eqnarray}\label{DensityDef}
D^{(n)}=\lim_{|A|\to\infty}\lim_{t\to\infty}\frac{S_A^{(n)}(t)}{|A|}
\end{eqnarray}
where the second limit is taken to avoid any boundary effects.  In \cite{AlbaCalabreseRenyi1, AlbaCalabreseRenyi2} it was shown that $D^{(n)}$ can be calculated for integrable models using a modification of the Quench Action formalism.  In particular, it can be written as 
\begin{eqnarray}\label{Density}
D^{(n)}&=&\int_0^\pi dk \, d^{(n)}(k)+\sum_{m=1}^\infty \int_0^\infty d\lambda \left[ d_m^{(n)}(\lambda)+ d'^{(n)}_m(\lambda)\right],
\end{eqnarray}
where the constituents are 
\begin{eqnarray}
d^{(n)}(k)&=&\frac{2(\rho(k)+\rho^h(k))}{1-n} \Bigg\{\log{\left[(1-\vartheta(k))^n+\frac{\vartheta(k)^n}{x^{(n)}(k)} \right]}+ \vartheta(k) \log{[x^{(n)}(k)]}\Bigg\},\nonumber\\
d^{(n)}_m(\lambda)&=&\frac{2(\sigma_m(\lambda)  +\sigma_m^h(\lambda) )}{1-n}\Bigg\{\log{\left[(1-\vartheta_m(\lambda))^n+\frac{\vartheta_m(\lambda)^n}{x^{(n)}_m(\lambda)} \right]}+ \vartheta_m(\lambda)  \log{[x^{(n)}_m(\lambda)]}\Bigg\},\label{eq:Renyidensity}\\\nonumber
d'^{(n)}_m(\lambda)&=&\frac{2(\sigma'_m(\lambda)  +\sigma_m^{\prime h}(\lambda) )}{1-n}\Bigg\{\log{\left[(1-\vartheta'_m(\lambda))^n+\frac{\vartheta'_m(\lambda)^n}{x^{\prime (n)}_m(\lambda)} \right]}+ \vartheta'_m(\lambda)  \log{[x^{\prime (n)}_m(\lambda)]}\Bigg\}.
\end{eqnarray}
We recall that here $\vartheta(k), \vartheta_m(\lambda), \vartheta'_m(\lambda)$ are the filling functions \eqref{eq:fillingfunctions} and we introduced the auxiliary functions $x^{(n)}(k)$, $x^{(n)}_m(\lambda)$, $x'^{(n)}_m(\lambda)$ which obey a set of TBA-like integral equations. In their partially decoupled form they read as  
\begin{eqnarray}\label{DensityTBA1}
\log x^{(n)}(k)&=&-s*\log\left[\left((1-\vartheta_1)^n x^{(n)}_1+ \vartheta_1^n\right)\left((1-\vartheta'_1)^n x^{\prime (n)}_1+ \vartheta_1^{\prime n}\right)\right](s_k)\\\nonumber
\log x^{(n)}_m(\lambda)&=&s*\log\left[\left((1-\vartheta_{m+1})^n x^{(n)}_{m+1}+ \vartheta_{m+1}^n\right)\left((1-\vartheta_{m-1})^n x^{(n)}_{m-1}+ \vartheta_{m-1}^{n}\right)\right](\lambda)\\\label{DensityTBA2}
&&-\int_{-\pi}^\pi dk\, s(\lambda-s_k)\log \left[(1-\vartheta(k))^n {x^{(n)}(k)} +{\vartheta(k))^n}\right]\\\nonumber
\log x'^{(n)}_m(\lambda) &=&s*\log\left[\left((1-\vartheta'_{m+1})^n x^{\prime (n)}_{m+1}+ \vartheta_{m+1}^{\prime n}\right)\left((1-\vartheta'_{m-1})^n x^{\prime (n)}_{m-1}+ \vartheta_{m-1}^{\prime n}\right)\right](\lambda)\\\label{DensityTBA3}
&&-\int_{-\pi}^\pi dk\, s(\lambda-s_k)\log \left[(1-\vartheta(k))^n {x^{(n)}(k)} +{\vartheta(k))^n}\right].
\end{eqnarray}
which can be numerically integrated in a manner similar to TBA equations. 

By exchanging the role of space and time in~\eqref{DensityDef} we can instead obtain the asymptotic slope of the entropy growth,
\begin{eqnarray}\label{SlopeDef}
S^{(n)}=\lim_{t\to\infty}\lim_{|A|\to\infty}\frac{S_A^{(n)}(t)}{2t}.
\end{eqnarray}
As shown in Ref.~\cite{bertini2022growth} this quantity can be related to $D^{(n)}$ via a space-time swap which allows to calculate it using the Quench Action. The result is also given in terms of a set of functions $y^{(n)}(k), y^{(n)}_m(\lambda),y'^{(n)}_m(\lambda)$,  and depends upon both the long time steady state as well as the quasi particle velocities $v(k), v_m(\lambda),~v'_m(\lambda)$ introduced in Sec.~\ref{sec:EE}.  Once again we write it as a sum over contributions of the particles, strings and bound states 
\begin{eqnarray}\label{Slope}
&&S^{(n)}=\int_0^\pi dk s^{(n)}(k)+\sum_{m=1}^\infty \int_0^\infty d\lambda \left[s^{(n)}_m(\lambda)+  s'^{(n)}_m(\lambda)\right]\,,
\end{eqnarray}
where we introduced 
\begin{eqnarray}
s^{(n)}(k)&=&\frac{2 v(k) (\rho(k)+\rho^h(k))}{1-n} \Bigg\{\log{\left[(1-\vartheta(k))^n+\frac{\vartheta(k)^n}{y^{(n)}(k)} \right]}+ \vartheta(k) \log{[y^{(n)}(k)]}\Bigg\},\\\nonumber
s^{(n)}_m(\lambda)&=&\frac{2 v_m(\lambda) (\sigma_m(\lambda)  +\sigma_m^h(\lambda) )}{1-n}\Bigg\{\log{\left[(1-\vartheta_m(\lambda))^n+\frac{\vartheta_m(\lambda)^n}{y^{(n)}_m(\lambda)} \right]}+ \vartheta_m(\lambda)  \log{[y^{(n)}_m(\lambda)]}\Bigg\},\\\nonumber
s'^{(n)}_m(\lambda)&=&\frac{2 v'_m(\lambda) (\sigma'_m(\lambda)  +\sigma_m^{\prime h}(\lambda) )}{1-n}\Bigg\{\log{\left[(1-\vartheta'_m(\lambda))^n+\frac{\vartheta'_m(\lambda)^n}{y^{\prime (n)}_m(\lambda)} \right]}+ \vartheta'_m(\lambda)  \log{[y^{\prime (n)}_m(\lambda)]}\Bigg\},
\end{eqnarray}
The auxiliary functions $y^{(n)}(k)$, $y^{(n)}_m(\lambda)$, $y'^{(n)}_m(\lambda)$ appearing in these expressions satisfy 
\begin{eqnarray}\label{SlopeTBA1}
\gamma(k) \log y^{(n)}(k)&=&-s*\gamma \log\left[\left((1-\vartheta_1)^n y^{(n)}_1+ \vartheta_1^n\right)\left((1-\vartheta'_1)^n y^{\prime (n)}_1+ \vartheta_1^{\prime n}\right)\right](s_k),\\\nonumber
\gamma(\lambda)\log y^{(n)}_m(\lambda)&=&s* \gamma \log\left[\left((1-\vartheta_{m+1})^n y^{(n)}_{m+1}+ \vartheta_{m+1}^n\right)\left((1-\vartheta_{m-1})^n y^{(n)}_{m-1}+ \vartheta_{m-1}^{n}\right)\right](\lambda)\\\label{SlopeTBA2}
&&-\int_{-\pi}^\pi dk\, s(\lambda-s_k)\gamma(k)\log \left[(1-\vartheta(k))^n {y^{(n)}(k)} +{\vartheta(k))^n}\right],\\\nonumber
\gamma(\lambda)\log y'^{(n)}_m(\lambda) &=&s*\gamma \log\left[\left((1-\vartheta'_{m+1})^n y^{\prime (n)}_{m+1}+ \vartheta_{m+1}^{\prime n}\right)\left((1-\vartheta'_{m-1})^n y^{\prime (n)}_{m-1}+ \vartheta_{m-1}^{\prime n}\right)\right](\lambda)\\\label{SlopeTBA3}
&&-\int_{-\pi}^\pi dk\, s(\lambda-s_k) \gamma(k) \log \left[(1-\vartheta(k))^n {y^{(n)}(k)} +{\vartheta(k))^n}\right],
\end{eqnarray}
where we set $\gamma(x)=\text{sgn}(x)$.  

\begin{figure}
 \centering
    (a)%
    \raisebox{-\totalheight+\baselineskip}[0pt][\totalheight]{\includegraphics[width=0.45\columnwidth]{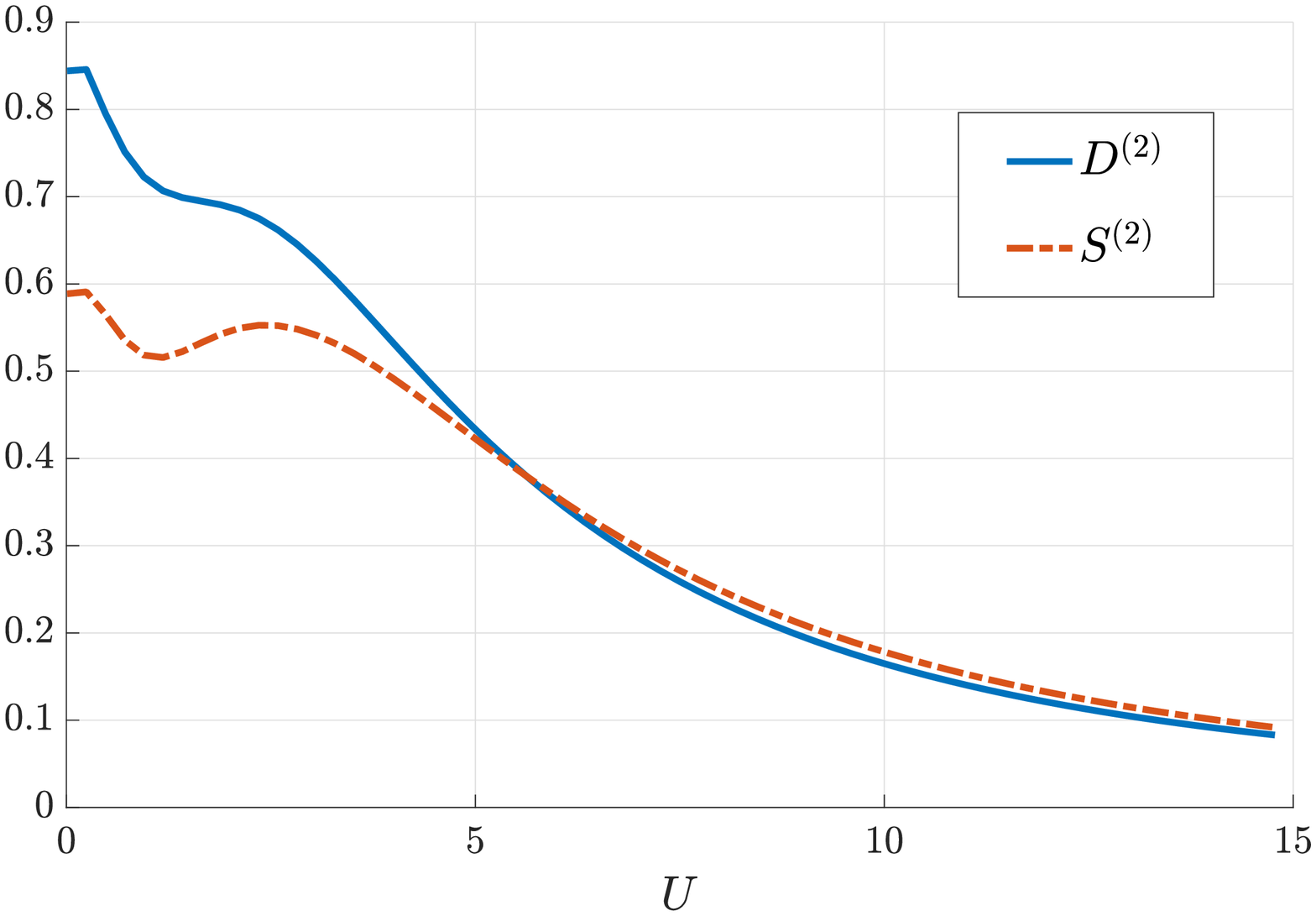}}\hfill
    (b)%
    \raisebox{-\totalheight+\baselineskip}[0pt][\totalheight]{\includegraphics[width=0.45\columnwidth]{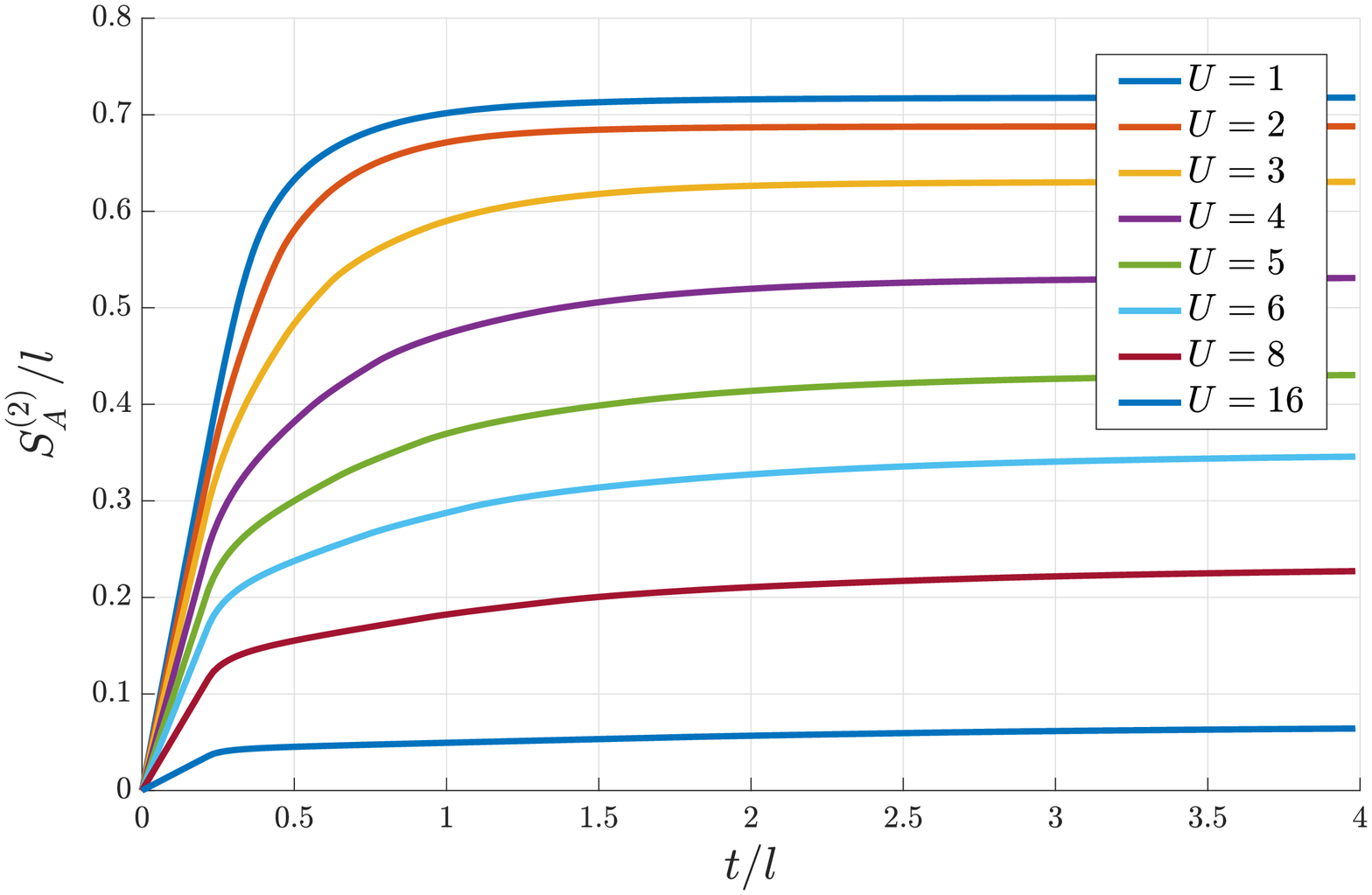}}
\caption{\label{fig:Renyi}
(a) The density~\eqref{DensityDef}, $D^{(2)}$ and slope~\eqref{SlopeDef}, $s_2$ of the R\'enyi-$2$ entanglement entropy as a function of $U$.  We take $\tilde{\alpha}=1$ which is close to half filling.  (b) The time evolution of the R\'enyi-2 entanglement entropy, $S^{(2)}_A/\ell$ as a function of $t/\ell$ for $U=1,\dots 8,16$.  
}
\end{figure}

We plot the results of the numerical evaluation of both the slope $S^{(2)}$ and density $D^{(2)}$ for $n=2$ as a function of the interaction strength in Fig.~\ref{fig:Renyi} (a).  As it has been found in previously studied cases, the two exhibit qualitatively similar behavior~\cite{bertini2022growth} and, as was the case for the von Neumann entropy, decrease with increasing $U$.  

A surprising consequence of Eq.~\eqref{Slope} is that the  quasiparticle picture does not apply to R\'enyi entropies in the presence of interactions. Namely, the quasi particle formula \eqref{EntEnt} with the Yang-Yang entropy densities \eqref{sec:yangyang} replaced by the R\'enyi entropy densities \eqref{eq:Renyidensity} does not describe the evolution of \eqref{eq:Renyi}~\cite{bertini2022growth, klobas2021entanglement}. To access the full evolution of the R\'enyi entropies Ref.~\cite{bertini2022growth} proposed an alternative formula inspired by \eqref{Slope} and tested it in the case of the quantum cellular automaton Rule 54.  For the Hubbard model the prediction, valid in the scaling limit, reads
\begin{eqnarray}\nonumber
S^{(n)}_A(t)&=&\int_{0}^\pi dk\, \text{min}\left( \ell d^{(n)}(k),2t s^{(n)}(k)\right)\\
&&+\sum_{n=1}^\infty\int_{0}^\infty\!\! d\lambda\,\Bigg\{ \text{min}\left(\ell d^{(n)}_m(\lambda),2ts^{(n)}_m(\lambda)\right)+\text{min}\left(\ell d'^{(n)}_m(\lambda),2t s'^{(n)}_m(\lambda)\right)\Bigg\}.
\end{eqnarray}
In the limit of $n\to 1$ this can be shown to agree with the quasiparticle picture given above but away from this limit has no interpretation in terms the dynamics of pairs of quasiparticles above the steady state.  We plot $S^{(2)}_A(t)$ as a function of time for different values of the interaction strength in Figure~\ref{fig:Renyi} (b). We see a similar structure to the von Neumann entanglement entropy including the emergence of the two distinct slopes around $U\approx 4$.

\section{Conclusions}
\label{SectionX}

In this paper we drew on previous work on integrable $SU(2|2)$-invariant spin chains~\cite{GB1,GB2} to identify a class of integrable initial states for the one-dimensional Hubbard model. We then proposed a formula (cf. Eq.~\eqref{eq:overlapfull}) describing the overlaps between these states and the Bethe states of the system, which we tested in the limits of low density and infinite repulsion. We used our formula to provide an exact characterisation, for arbitrary values of the interaction, of the statistical ensemble describing local subsystems at long times after quenches from our integrable initial states. We also described the growth of entanglement using the quasiparticle picture for the von Neumann entropy~\cite{AlbaCalabrese1}, and the recently developed spacetime swap formalism for R\'enyi entropies~\cite{bertini2022growth}. The calculation of other observables (e.g., one- or two-point functions of local operators) requires the knowledge of the the corresponding form factors, as done for instance in the XXZ spin-chain~\cite{mestyan2014short}, the sinh-Gordon field theory~\cite{negro2013on, negro2014on, bertini2016quantum}, and the Lieb-Liniger model~\cite{pozsgay2011local, bastianello2018exact, bastianello2018from}. Although these are not yet known we hope that our results will spark further research in this direction.

Here we identified several different families of integrable initial states for the Hubbard model, but we only computed the overlap for a specific one. An interesting future direction would be to find analogous overlap formulas for different classes of states. For instance, this would allow for the exact characterisation of inhomogeneous quenches in the genuine out-of-equilibrium scenario of different non-stationary states joined together (along the lines of the ``global" inhomogeneous quenches studied in Ref.~\cite{BertiniColluraDenardisFagotti} for the XXZ spin chain). Furthermore, another class of treatable initial states are integrable matrix product states that could be constructed along the lines of Ref. \cite{PiroliPozsgayVernier2} for the XXZ spin-chain. 

Another interesting direction, which we undertake in a companion paper Ref.~\cite{rylands2022wip}, is to use our formula to study the quenches in the Gaudin-Yang electron gas via a continuum limit. The latter system is particularly interesting in view of its direct experimental accessibility. Finally, recent experiments with ultra-cold alkaline-earth atoms~\cite{pagano2014one} motivate the study of Hubbard-like models (and their continuum counterparts) with $SU(N)$ internal symmetry. For the quench problem, such study should be possible by some non-trivial adaptation of our technique to obtain the overlaps. 

\acknowledgements

We thank Tamas Gombor for drawing our attention to Refs.~\cite{GB1, GB2}. BB was supported by the Royal Society through the University Research Fellowship No. 201102.
PC and CR acknowledge support from the ERC under Consolidator grant number 771536 (NEMO).  

\appendix

\section{Further partially decoupled equations}
\label{app:decoupledequations}

Here we report the explicit, partially decoupled form of some of the TBA equations used in the main text. 

\subsection{Partially decoupled form of Eqs.~(\ref{BetheHellFeyn1})--(\ref{BetheHellFeyn3})}

The partially decoupled form of Eqs.~(\ref{BetheHellFeyn1})--(\ref{BetheHellFeyn3}) reads as
\begin{eqnarray}\nonumber
[1+\eta_n(\lambda)]\mu_n(\lambda)&=&s*[\eta_{n+1}\mu_{n+1}+\eta_{n-1}\mu_{n-1}](\lambda)+\delta_{n,1}\int_{-\pi}^\pi {\rm d}k s(s_k-\lambda)\left[\cos{(k)}\omega(k)\right]\\\label{DecoupledBetheHellFeyn1}
&&+f*\left[\sigma_n(\lambda)+\sigma_n^h(\lambda)\right]\\ \nonumber
\left[1+\eta'_n(\lambda)\right] \mu'_n(\lambda)&=&s*[\eta'_{n+1}\mu'_{n+1}+\eta'_{n-1}\mu'_{n-1}](\lambda)-\delta_{n,1}\int_{-\pi}^\pi {\rm d}k s(s_k-\lambda)\left[\cos{(k)}\omega(k)\right]\\\label{DecoupledBetheHellFeyn2}
&&+f*[{\sigma'}_n(\lambda)+{\sigma'}_n^h(\lambda)]\\\label{DecoupledBetheHellFeyn3}
\left[1+\zeta(k)\right]\omega(k)&=& -s*[\eta_1\mu_1+\eta'_1\mu'_1](s_k)+f*[\sigma_1^h(s_k)+{\sigma'_1}^h(s_k)]-s*[f*\frac{1}{2\pi}{\dot{p}_{k-\lambda,1}}+\frac{1}{2\pi}{{p'}_{k-\lambda,1}}]
\end{eqnarray}
where used the shorthand notation \eqref{eq:shorthand} and we introduced 
\be
f(x)=(2U)^{-1}\text{csch}[\pi x/U].
\ee  
and where $\dot{p}_{k-\lambda,n}=\partial_x p_{k-\lambda,n}(x),~{p'}_{k-\lambda,n}=\partial_U p_{k-\lambda,n}(x)$

\subsection{Partially decoupled form of the dressed-velocity equations}

The partially decoupled form of the dressed-velocity equations read as 
\begin{eqnarray}
\nonumber
[1+\eta_n(\lambda)]\sigma_n(\lambda)v_n(\lambda)&=& s*\left[\eta_{n+1}\sigma_{n+1}v_{n+1}+\eta_{n-1}\sigma_{n-1}v_{n-1}\right](\lambda)+\delta_{n,1}\int_{-\pi}^\pi {\rm d}k s(s_k-\lambda)\rho(k)v(k)\\\nonumber
\left[1+\eta'_n(\lambda)\right] \sigma'_n(\lambda)v'_n(\lambda)&=&s*[\eta'_{n+1}\sigma'_{n+1}v_{n+1}+\eta'_{n-1}\sigma'_{n-1}v_{n-1}](\lambda)-\delta_{n,1}\int_{-\pi}^\pi {\rm d}k s(s_k-\lambda)\rho(k)v(k)\\\nonumber
&&+\delta_{n,1}\,t\int_{-\pi}^\pi\frac{{\rm d}k}{8\pi^2}\cos^2{(k)}\dot{s}(\lambda-s_k)
\\\nonumber
\left[1+\zeta(k)\right]\rho(k)v(k)&=&\frac{t\sin(k)}{\pi}+\cos{(k)}\int_{-\infty}^\infty {\rm d}\lambda\Big\{ \frac{1}{2\pi} \phi_1(s_k-\lambda)\sigma_0(\lambda)\\
&&-s(s_k-\lambda)(\eta_1(\lambda)\sigma_1(\lambda)v_1(\lambda)+\eta'_1(\lambda)\sigma'_1(\lambda)v'_1(\lambda))\Big\}.
\end{eqnarray}
where $\dot{s}=\partial_x s(x)$ and we have used~\eqref{IntEnergy}.

\bibliographystyle{apsrev4-1}

\bibliography{Hubbardbib.bib}

\end{document}